\documentclass[12pt]{article}

\pdfoutput=1

\usepackage{placeins}
\usepackage{amsmath,amssymb,amsfonts,amsthm,bm}    
\usepackage{algorithm2e}
\usepackage{graphicx}
\usepackage{lineno}
\usepackage{pdfpages}
\usepackage[onehalfspacing]{setspace}
\usepackage[margin=1.5in]{geometry}
\usepackage{subcaption}
\usepackage{setspace}
\usepackage{placeins}
\usepackage{url}
\usepackage{xcolor}
\usepackage{mathtools}
\usepackage{titlesec}
\usepackage{natbib}
\usepackage{fixmath}
\usepackage{bbm}
\usepackage{booktabs}
\usepackage{multirow,bigdelim}
\usepackage{comment}
\usepackage{commath}

\titlelabel{\thetitle.\quad}

\newcommand\sbullet[1][.5]{\mathbin{\vcenter{\hbox{\scalebox{#1}{$\bullet$}}}}}

\begin{document}
	
\def\spacingset#1{\renewcommand{\baselinestretch}%
{#1}\small\normalsize} \spacingset{1}
	

\title{\bf Estimation of Latent Network Flows in Bike-Sharing Systems}
\author{Marc Schneble and G\"oran Kauermann
\hspace{.2cm}\\
Department of Statistics, Ludwig-Maximilians Universit\"at M\"unchen}
		
\date{}	
		
\maketitle

\bigskip
\begin{abstract}
\noindent 
Estimation of latent network flows is a common problem in statistical network analysis. The typical setting is that we know the margins of the network, i.e. in- and outdegrees, but the flows are unobserved. In this paper, we develop a mixed regression model to estimate network flows in a bike-sharing network if only the hourly differences of in- and outdegrees at bike stations are known. We also include exogenous covariates such as weather conditions. Two different parameterizations of the model are considered to estimate 1) the whole network flow and 2) the network margins only. The estimation of the model parameters is proposed via an iterative penalized maximum likelihood approach.  This is exemplified by modeling network flows in the Vienna Bike-Sharing Network. Furthermore, a simulation study is conducted to show the performance of the model. For practical purposes it is crucial to predict when and at which station there is a lack or an excess of bikes. For this application, our model shows to be well suited by providing quite accurate predictions.
\end{abstract}
	
\noindent%
\vfill

\section{Introduction}

In many urban areas bike-sharing systems offer the possibility to rent bicycles on a short-term basis. Typically, the bikes can be rented and returned from stations that are distributed within a service region. In June 2019, about two thousand bike-sharing systems had been established around the world -- most of them in Europe, Eastern Asia and in Northern America. Another hundreds of them are in planning or under construction\footnote{\url{http://www.bikesharingmap.com}}. The growing popularity can be explained by several factors. The increasing use of public transportation systems led to a demand for the solution of the ``first/last mile problem'' (e.g. \citeauthor{shaheen2010bikesharing}, \citeyear{shaheen2010bikesharing}) for that a reasonable allocated bike-sharing system is a possible answer. Even more, bike-sharing systems itself are a fast and cheap way to implement urban transportation systems to travel distances of typically 1-5 km for a decent price (\citeauthor{midgley2011bicycle}, \citeyear{midgley2011bicycle}).   

Without interventions of the providers, a permanent imbalance of the station feeds would occur, meaning that some stations do not have available bikes while others have an overflow. Therefore, being able to estimate bicycle usage in a system offers the possibility to rebalance bikes before these imbalances arise. However, only few providers such as ``Citi Bike'' in New York City or ``Divvy'' in Chicago offer public available complete trip data\footnote{\url{https://www.citibikenyc.com/system-data}; \url{https://www.divvybikes.com/system-data}}. Whenever we refer to complete trip data, this means that at least the departure time and station as well as the destination time and station are known for every trip in the observation period. With data of Citi Bike NYC, \cite{li2015traffic} clustered bike stations into groups in order to implement a hierarchical prediction model to predict the number of bikes that will be rented from/returned to each station. Using the Divvy data from Chicago, \cite{zhang2016bicycle} build a trip destination inference model as well as a trip duration inference model. 

If complete trip data are not available (e.g. due to data confidentiality issues) one can often at least gather real-time information on stations feeds. Using station feed data for every minute over a period of two years, \cite{chen2017understanding} inferred the hourly incoming and outgoing traffic for every station. Since only a fraction of the rental and return events occur within the same minute, the error when compared to the complete flow data was only 0.05\%. Here, the trip inference problem was transformed to an ill-posed linear inverse problem (\citeauthor{airoldi2013estimating}, \citeyear{airoldi2013estimating}). To overcome this problem, the authors used a mixture of a Ridge and a Lasso approach. However, some disadvantages of this model are that their results are hardly interpretable and the need for minute wise data.

In this paper, we consider station feed data that are available for 1 hour intervals. Unlike the minute wise data, we can not in this case infer the incoming and outgoing traffic with the same high precision since there are many rental and return events occurring in the same time interval. However, we can observe the hourly differences of station feeds over time. Hence, for every time interval we aim to estimate the network flow with a cardinality of $N^2$ possible connections observing only $N$ differences of station feeds.

We propose a regression model that takes additional exogenous covariates into account such as weather data and the geographic coordinates of the stations. The network flows are modeled as independently Poisson-distributed which induces that the differences of the station feeds are Skellam-distributed (\citeauthor{skellam1948probability}, \citeyear{skellam1948probability}). Similar modeling approaches were proposed by \cite{karlis2008bayesian} for the modeling of the goal difference in football games and by \cite{koopman2014dynamic} to investigate high-frequency returns in trading. The use of a Skellam distribution in the field of network analysis has been proposed in \cite{gan2018approximation}. The authors represent the difference of the number of edges between two graphs via an approximation of Skellam distributed random variables. Further applications of the Skellam distribution are amongst others concerned with the measuring of the intensity difference of pixels in the spatial and temporal domain (\citeauthor{hwang2007sensor}, \citeyear{hwang2007sensor}) and the activation of neurons related with finger movements (\citeauthor{shin2010neural}, \citeyear{shin2010neural}).

This paper is organized as follows. In Section \ref{sec: model and notation} we introduce the regression model and the notation for our analyses. Subsequently in Section \ref{sec: model estimation} we show how to estimate the model parameters. In Section \ref{sec: application vienna} we are concerned with the application to the Vienna Bike-Sharing System including the results and a model evaluation. The main part of this paper is completed by a simulation study in section \ref{sec: simulation}. In Section \ref{sec: discussion}, we briefly discuss the main results of this paper.

\section{Model and Notation}
\label{sec: model and notation}


\subsection{Poisson Modeling of Trip Counts}

In our analysis we consider a temporal network having $N$ nodes (stations) and therefore $N^2$ possible edges (routes between stations), where we also allow for self-loops. For the discrete sequence of points in time  $t = 0,1,\dots,T$ we observe a realization of the $\mathbb{N}_0$-valued random variable $\mathcal{C}_{i,t}$ (station feeds) on every node $i = 1,\dots,N$. We denote with $\mathcal{N}_{ij,t}$ the count of trips from station $i$ to station $j$ departing in the interval $[t-1,t)$ and choose each time interval to be one hour. Our aim is to estimate the network flows $\mathcal{N}_{ij,t}$ based on the hourly station feeds $\mathcal{C}_{i,t}$. The counts of trips are modeled separately for each hour of the day and with $\mathcal{T}$ we denote the corresponding set of points in time. Hence, $\mathcal{T} = \lbrace h, h+24, h+48, \dots, T-24+h \rbrace$ for some $h \in \lbrace 1,\dots,24 \rbrace$. We start by assuming a log-linear Poisson model for the trip counts $\mathcal{N}_{ij,t} \sim \text{Poi}(\mu_{ij,t})$ where
\begin{align}
\label{eq: pois_model}
\mu_{ij,t} = \exp(\eta_{ij,t}) =  \exp\left( \eta(\mathbold{z}_{ij,t}) +  u_i^{\text{out}}  +u_j^{\text{in}}  \right).
\end{align}
With $\mathbold{z}_{ij,t}$ we denote covariates which may be dyadic and time specific (i.e. $\mathbold{z}_{ij,t} = \mathbold{x}_{ij,t}$), dyadic specific only (i.e. $\mathbold{z}_{ij,t} = \mathbold{x}_{ij}$), station (and time) specific (i.e. $\mathbold{z}_{ij,t} = \mathbold{x}_{i,t}$ for outgoing or $\mathbold{z}_{ij,t} = \mathbold{x}_{ j,t}$ for ingoing) or just time specific (i.e. $\mathbold{z}_{ij,t} = \mathbold{x}_t$). To keep the notation general we will, wherever possible, denote these with all three indices.

The coefficients $u_i^\text{out}$ and $u_j^\text{in}$ are station specific out- and indegree effects which we specify as random effects that account for unobserved station specific heterogeneity. The random effects $\mathbold{u}_i$ are modeled as independently bivariate normally distributed, i.e. 
\begin{equation}
\mathbf{u}_i = \left(u_i^{\text{out}}, u_i^{\text{in}}\right)^\top \sim \mathcal{N} \left( \mathbf{0}, \mathbf{\Sigma} \right)
\label{eq: distribution_random_effects}
\end{equation}
where $\mathbf{\Sigma}$ is the variance matrix which needs to be estimated from the data.  

The fixed covariate effects $\eta(\mathbold{z}_{ij,t})$ are modeled parametrically as well as semiparametrically using penalized splines, see \cite{eilers1996flexible}, \cite{ruppert2003semiparametric} or \cite{fahrmeir2007regression}. To be specific, the linear predictor $\eta(\mathbold{z}_{ij,t})$ is constructed from 
\begin{equation}
\eta(\mathbold{z}_{ij,t}) = \mathbold{z}_{ij,t}^\text{lin}\mathbold{\beta} + \sum_{m=1}^M s_m(z_{ij,t}^{(m)}).
\label{eq: z_ijt}
\end{equation}
where the row vector $\mathbold{z}_{ij,t} = \left(\mathbold{z}_{ij,t}^{\text{lin}}, z_{ij,t}^{(1)}, \dots, z_{ij,t}^{(M)}\right)$ consists of station-, route- and time-specific covariate values concerning trips from station $i$ to station $j$ departing in the interval $[t-1, t)$. The row vector $\mathbold{z}_{ij,t}^{\text{lin}}$ contains covariates modeled linearly and $\mathbold{\beta}$ is the vector of the corresponding parameters including an intercept. The scalars $z_{ij,t}^{(m)}$ represent effects that are modeled semiparametrically and $s_m(\cdot)$ are smooth functions in $z_{ij,t}^{(m)}$. We represent $s_m(\cdot)$ through a basis representation
\begin{equation*}
s_m(z_{ij,t}^{(m)}) = \sum_{r=1}^{k_m} B_m^{(r)}(z_{ij,t}^{(m)}) \mathbold{\gamma}_m^{(r)}
\end{equation*}
where $\mathbold{\gamma}_m \in \mathbb{R}^{k_m}$ is a vector of basis coefficients and $\mathbold{B}_m(\cdot) = \left( B_m^{(1)}(\cdot),\dots,B_{m}^{(k_m)}(\cdot) \right)^\top$ is a B-spline basis function constructed on knots $\tau_1,\dots,\tau_{k_m}$. For the seasonal effect, we use cyclic splines to ensure annual continuity. To achieve identifiability of the $M$ spline functions, we enforce for every $s_m(\cdot)$ that the function integrates out to zero. In practice, this is enforced by setting the empirical function mean to zero which can be implemented by centering the columns of $\mathbold{B}_m$ around zero. According to \cite{wood2017generalized}, we specify an improper normal prior  with variance matrix $\sigma_m^2 \mathbold{K}_m^{-}$ on the spline-parameters $\mathbold{\gamma}_m$ where $\mathbold{K}_m$ serves as a penalty matrix which is constructed from second-order differences and $\mathbold{K}_m^{-}$ denotes the generalized inverse of $\mathbold{K}_m$. This setting allows to estimate the $M$ smoothing parameters $\lambda_m = \frac{1}{\sigma^2_m}$ in the process of estimating $\mathbf{\Sigma}$.

\subsection{Dyadic Modeling of Trip Counts}

A bike trip from station $i$ to station $j$ which departs by our definition in the interval $[t-1,t)$ does not need to reach its destination within the same time interval. Instead, a customer could also arrive in the subsequent interval $[t, t+1)$. We additionally account for these trips by installing an additional latent station, denoted by $w$. Hence, for every time point $t$, each bike is either parked in one of the $N$ physical stations or it is on the way, which we formally model as being allocated to the latent station $w$. We assume that trips do not last for more than two time intervals, which is reasonable for the large majority of trips. In other words, we do not allow self-loops for the latent station $w$. 

The modeling approach described above does not allow to model the trip counts $\mathcal{N}_{ij,t}$ directly by exploiting the station feeds $\mathcal{C}_{i,t}$ only since trips not departing and arriving in the same hour are ignored. We therefore need to change notation and define with $\mathcal{Y}_{ij,t}$ the count of trips from station $i$ to station $j$ departing and ending in the time interval $[t-1,t)$. Accordingly, $\mathcal{Y}_{iw,t}$ and $\mathcal{Y}_{wj,t}$ are trips not starting and ending in the same time interval $[t-1,t)$. The first denotes the count of trips that start in the current time interval having station $i$ as origin where the actual destination remains unspecified. Likewise, $\mathcal{Y}_{wj,t}$ are the trips that started in the previous time interval at an unknown origin and end at $j$. We assume the just defined trip counts to be Poisson distributed, i.e. $\mathcal{Y}_{ij,t} \sim \text{Poi}(\nu_{ij,t})$ for $(i,j) \in \lbrace 1,\dots,N,w \rbrace^2 \backslash \lbrace (w,w) \rbrace$ and $t\in\mathcal{T}$ where
\begin{align}
\label{eq: nu}
\nu_{ij,t} = \exp(\eta_{ij,t}) =  \exp\left( \eta(\mathbold{z}_{ij,t}) +  u_i^{\text{out}}  +u_j^{\text{in}}  \right)
\end{align}
is defined according to the Poisson modeling approach proposed in the previous section.

If we assume the trip counts $\mathcal{Y}_{ij,t}$ to be independent given the covariates and random effects, the counts of outgoing bikes $\mathcal{N}_{i \sbullet,t }$ \textit{from} station $i$ and the counts of incoming bikes $\mathcal{N}_{\sbullet i, t}$ \textit{to} station $i$, respectively, are again Poisson-distributed, so that
\begin{align}
\label{eq: mu_i_dot_t}
\mathcal{N}_{i \sbullet ,t} &= \sum_{j=1}^N \mathcal{Y}_{ij,t} + \mathcal{Y}_{iw,t} \sim \text{Poi}\left( \sum_{j=1}^N {\nu}_{ij,t} + \nu_{iw,t} \right) = \text{Poi}(\mu_{i \sbullet ,t}), \\
 \label{eq: mu_dot_j_t}
\mathcal{N}_{\sbullet i, t} &= \sum_{j=1}^N \mathcal{Y}_{ji,t} + \mathcal{Y}_{wi,t} \sim \text{Poi}\left( \sum_{j=1}^N {\nu}_{ji,t} + \nu_{wi,t} \right) = \text{Poi}(\mu_{\sbullet i,t}).
\end{align}
With $\mathcal{C}_{i, t-1}$ as station count of the $i$-th station in $t-1$ we obtain  $\mathcal{C}_{i, t} = \mathcal{C}_{i, t-1} + \mathcal{N}_{\sbullet i, t} - \mathcal{N}_{i \sbullet, t}$. Defining with 
\begin{align*}
\mathcal{D}_{i, t} = \mathcal{C}_{i, t} - \mathcal{C}_{i, t-1} = \left( \sum_{j=1}^N \mathcal{Y}_{ji, t} + \mathcal{Y}_{wi, t} \right) - \left( \sum_{j=1}^N  \mathcal{Y}_{ij, t} + \mathcal{Y}_{iw, t}  \right) 
\end{align*}
 the difference in the $i$-th station count from $t-1$ to $t$, we obtain for $\mathcal{D}_{i, t}$ a Skellam distribution with parameters $\mu_{\sbullet i, t}$ and $\mu_{i \sbullet, t}$, see e.g. \cite{alzaid2010poisson}. More precisely, if $X \sim \text{Poi}(\theta_1)$ and $Y \sim \text{Poi}(\theta_2)$ are independent, then $D = X - Y \sim \text{Skellam}(\theta_1, \theta_2)$ and the probability mass function of $D$  is given by 
\begin{align}
\notag
\mathbb{P}(D = d) = \exp \left( -\theta_1 - \theta_2 \right) \left( \frac{\theta_1}{\theta_2} \right)^\frac{d}{2} I_{|d|}\left(2 \sqrt{\theta_1 \theta_2}\right)
\intertext{for $d \in \mathbb{Z}$ where}
I_{d}(\theta) =  \left( \frac{\theta}{2} \right)^d \sum_{k=0}^\infty \left( \frac{\theta}{2} \right)^{2k} \frac{1}{k!(d+k)!} 
\label{eq: modified bessel function}
\end{align}
is the modified Bessel function of the first kind (\citeauthor{abramowitz1965handbook}, \citeyear{abramowitz1965handbook}). A ratio test can be applied to show the absolute convergence of the series in \eqref{eq: modified bessel function}. However, this series does not need to converge numerically. If this is the case, we compute the logarithms and the ratios of the modified Bessel function, which we need to fit the model, making use of approximations developed by \cite{amos1974computation}. Details are given in Appendix \ref{app: skellam}. Furthermore, we denote with $l_D(\mathbold{\theta}; d) = \log \mathbb{P}_\mathbold{\theta}(D = d)$ the log-likelihood contribution. The derivatives of the log-likelihood, which are required to fit our model, are also elaborated in Appendix \ref{app: skellam}.

If we assume that the total count of bikes in the system at time $t-1$ equals the count at $t$, then
\begin{equation*}
\mathcal{D}_{w, t} = \sum_{j=1}^N \mathcal{Y}_{jw, t} - \sum_{j=1}^N \mathcal{Y}_{wj, t}  = - \sum_{i=1}^N \mathcal{D}_{i, t},
\end{equation*} 
i.e. the differences of the physical station feeds imply the differences of the latent station's feeds which are again Skellam-distributed. Thus, our regression model results to 
\begin{equation*} 
\mathcal{D}_{i, t} \sim \text{Skellam}(\mu_{\sbullet i, t}, \mu_{i \sbullet, t}) \label{eq: Model}
\end{equation*}
for $i \in  \lbrace 1,\dots,N,w  \rbrace$ and $t \in \mathcal{T}$ where $\mu_{\sbullet i, t}$ and $ \mu_{i \sbullet, t}$ are defined as in \eqref{eq: mu_i_dot_t} and \eqref{eq: mu_dot_j_t}, respectively, and $\mathcal{T}$ denotes the set of time points which belong to the evaluated hour of the day.

By estimating the model parameters (as shown in the next section) and inserting them into \eqref{eq: nu}, we get estimates $\nu_{ij,t}$ for $(i,j) \in \lbrace 1,\dots,N,w \rbrace^2 \backslash \lbrace(w,w)\rbrace$ and $t\in\mathcal{T}$. However, we actually want to estimate the expected trip counts $\mu_{ij,t}$ that were defined in the previous subsection. In other words, we also want to allocate (i.e. estimate) the destination for trips that start in one period but end in the subsequent period. For this, we merely need to estimate the probability $\pi_{ij,t}$ that a trip originating at station $i$ in $[t-1,t)$ which is exceeding this time interval, will terminate at station $j$. If the decision of the terminal station is independent of exceeding or not exceeding this time interval, which is a plausible assumption, this probability can be estimated by 
\begin{equation*}
\widehat{\pi}_{ij,t} = \frac{\widehat{\nu}_{ij,t}}{\sum_{j=1}^N \widehat{\nu}_{ij,t}}
\end{equation*}
and thus we set $\widehat{\mu}_{ij,t} = \widehat{\nu}_{ij,t} + \widehat{\nu}_{iw,t} \widehat{\pi}_{ij,t}$. This yields the final estimate for dyadic movements in the network based on station feeds.

\subsection{Station based Modeling of Trip Counts}
\label{subsec: station based model}

Amongst others, the above model builds on dyadic covariates $\mathbold{z}_{ij,t}$, i.e. quantities that are specific for a trip from $i$ to $j$. If covariates are available on a station level only,  we find $\mathbold{z}_{ij,t}$ to depend either on $i$ or on $j$ but not on both. In this case we can simplify the model since \eqref{eq: z_ijt} decomposes to 
\begin{equation*}
\eta(\mathbold{z}_{ij,t}) = \eta_\text{out}(\mathbold{x}_{i,t}) + \eta_{in}(\mathbold{x}_{j,t}) + \eta(\mathbold{x}_t) + u_i^\text{out} + u_j^\text{in}
\end{equation*}
where, as introduced before, $\mathbold{x}_{i,t}$ are outgoing specific covariates, $\mathbold{x}_{j,t}$ are ingoing specific and $\mathbold{x}_t$ is just time specific. It is easy to see that $\mu_{i \sbullet, t}$ simplifies to 
\begin{align*}
\mu_{i \sbullet,t} &= \exp \left( \eta_\text{out}(\mathbold{x}_{i,t}) + \widetilde{\eta}(\mathbold{x}_t) + u_i^\text{out}	\right)
\intertext{where}
\widetilde{\eta}(\mathbold{x}_t) &= \eta(\mathbold{x}_t) + \log  \left( \sum_{j=1}^N \exp (\eta_\text{in}(\mathbold{x}_{j,t}) + u_j^\text{in}) \right).
\end{align*}
Similarly, we obtain simplifications for $\mu_{\sbullet j,t}$. This model does not rely on the conditional independence assumption of the counts of imcoming/outgoing trips to/from a station any more. In the same way as in the dyadic modeling approach we can model the incoming and the outgoing trips to the latent station $w$.

The performance of both, the dyadic model and the station based model is limited by various characteristics of the network. Amongst others, the provider transports bikes between stations to work against imbalances. Moreover, broken bikes are taken out of the system and brought back after repair. Such information is not provided and accounts for inevitable inaccuracy of our models.

\section{Model Estimation}
\label{sec: model estimation}

Estimation of the model parameters is performed iteratively by an approximate EM-algorithm comparable to \cite{fahrmeir2001multivariate} in the setting of a generalized linear mixed model. The fundamental idea is to alternately estimate the model parameters and random effects $\mathbold{\theta} = \left(\mathbold{\beta}^\top, \mathbold{\gamma}_1^\top, \dots, \mathbold{\gamma}_M^\top, \mathbold{u}_1^\top, \dots, \mathbold{u}_N^\top, \mathbold{u}_w^\top\right)^\top$ as well as the variance components $\mathbf{\Sigma}$ and $\mathbold{\lambda} = (\lambda_1,\dots,\lambda_M)^\top$. Algorithm \ref{alg: EM} illustrates the procedure applied to our model.

\begin{algorithm}[h]
\SetAlgoLined
\KwResult{Estimates $\widehat{\mathbold{\theta}}, \widehat{\mathbf{\Sigma}}$ and $\widehat{\mathbold{\lambda}}$}
 Initialize starting values $\widehat{\mathbf{\Sigma}}^{(0)}$ and $\widehat{\mathbold{\lambda}}^{(0)}$;
 Set value for $\epsilon$; $p = 0$; 
 
 \textit{convergence} = \textit{FALSE};\
 
 \While{convergence = FALSE}{
  estimate $\mathbold{\theta}$ given $\widehat{\mathbf{\Sigma}}^{(p)}$ and $\widehat{\mathbold{\lambda}}^{(p)}$\;
  determine $\widehat{\mathbf{\Sigma}}^{(p+1)}$ and $\widehat{\mathbold{\lambda}}^{(p+1)}$\;
  \eIf{$|| \widehat{\mathbf{\Sigma}}^{(p+1)}- \widehat{\mathbf{\Sigma}}^{(p)}|| / || \widehat{\mathbf{\Sigma}}^{(p)} || < \epsilon$}{
   convergence = \textit{TRUE}\;
   }{
   $p = p+1$\;
  }
 }
 \caption{Approximate EM-algorithm}
 \label{alg: EM}
\end{algorithm}

In order to estimate $\mathbold{\theta}$, we make use of a Laplace approximation as generally proposed in \cite{breslow1993approximate}. It can be shown that this is equivalent to maximizing the penalized log-likelihood
\begin{equation*}
l_P(\mathbold{\theta}) =  \sum_{i \in \lbrace 1, \dots, N, w \rbrace} \sum_{t \in \mathcal{T}} l_D(\mathbold{\theta}; d_{i, t}) - \frac{1}{2} \sum_{m=1}^M \lambda_m \mathbold{\gamma}_m^\top \mathbf{K}_m \mathbold{\gamma}_m - \frac{1}{2}  \sum_{i \in \lbrace 1, \dots, N, w \rbrace} \mathbold{u}_i^\top \mathbf{\Sigma}^{-1} \mathbold{u}_i
\end{equation*}
where $l_D(\mathbold{\theta}; d_{i,t})$ is the log-likelihood contribution from above evaluated for an observation $D_{i,t} = d_{i,t}$ as well as $\mathbf{\Sigma}$ and $\mathbold{\lambda}$ being fixed to some value. The first penalty refers to the spline functions, the second to the random effects. 

The estimation of the covariance matrix $\mathbf{\Sigma}$ is then based on maximization of the resulting Laplace approximation. This in fact is a posterior mode estimation, but since posterior mode and posterior mean are close, \cite{fahrmeir2001multivariate} name the algorithm ``approximate EM-algorithm''. Formally, using the current estimate $\widehat{\mathbold{\theta}}$, the update of $\widehat{\mathbf{\Sigma}}$ is carried out through
\begin{align}
\label{eq: updata_Sigma}
\widehat{\mathbf{\Sigma}}^{(p+1)} &= \frac{1}{N+1}  \sum_{i \in \lbrace 1, \dots, N, w \rbrace} \left( \widehat{\mathbold{V}}_{u_i u_i} + \widehat{\mathbold{u}}_i \widehat{\mathbold{u}}_i^\top \right)
\end{align}
where $\widehat{\mathbold{V}} = (\mathbold{F}^{\text{obs}})^{-1}(\widehat{\mathbold{\delta}})$ is the inverse of the observed Fisher matrix of $\mathbold{\theta}$ and $\widehat{\mathbold{V}}_{u_i u_i}$ denotes the diagonal elements of $\widehat{\mathbold{V}}$ related to $\mathbold{u}_i$. 

In order to compute the update of $\widehat{\mathbold{\lambda}}$, we follow \cite{wood2017generalizedfellner}. Given the estimate $\widehat{\lambda}_m^{(p)}$ of the previous cycle and the current estimate $\widehat{\mathbold{\theta}}$, the updates are carried out through
\begin{equation}
\widehat{\lambda}_m^{(p+1)} = \frac{ \texttt{tr}(\mathbold{S}_\lambda^- \mathbold{S}_m)- \texttt{tr}(\widehat{\mathbold{V}} \mathbf{S}_m)}{\widehat{\mathbold{\theta}}^\top \mathbold{S}_m \widehat{\mathbold{\theta}}}\widehat{\lambda}_m^{(p)}
\label{eq: update_lambda}
\end{equation}
for $m = 1,\dots,M$. Here, $\mathbold{S}_m$ is the matrix $\mathbold{K}_m$ augmented with zeroes such that it fits the dimension of $\widehat{\mathbold{V}}$ and the entries of $\mathbold{K}_m$ in $\mathbold{S}_m$ are located at the same place as the entries of $\widehat{\mathbold{V}}_{\gamma_m \gamma_m}$ in $\widehat{\mathbold{V}}$, see also Appendix \ref{app: algorithm}. Furthermore,
\begin{equation*}
\mathbold{S}_\lambda = \sum_{m=1}^M \lambda_m^{(p)} \mathbold{S}_m
\end{equation*}
and \texttt{tr} denotes the trace operator for diagonal matrices. \cite{wood2017generalizedfellner} also show that if $\widehat{\mathbold{V}}$ is positive definite, the difference in the nominator of \eqref{eq: update_lambda} is guaranteed to be positive and hence $\widehat{\lambda}_m^{(p+1)}$, too. Since we do not use the expected Hessian but the observed Hessian, this is not necessarily fulfilled. In the case of a negative $\widehat{\lambda}_m^{p+1}$, they propose to replace it by a ``suitable nearest positive definite matrix to the observed Hessian''.

\section{Application to the Vienna Bike-Sharing Network}
\label{sec: application vienna}

We now apply our model to the Vienna Bike-Sharing Network with data from the year 2014. In this year, the Vienna Bike-Sharing System\footnote{\url{https://www.citybikewien.at/en/}} consisted of $N = 120$ stations whereby two of them were installed in the course of the year. According to \cite{BSAtlas}, it belongs to the 50 largest bike-sharing networks of the world. Besides the station feed data $C_{i,t}$ that induce the differences of station feeds $D_{i,t}$ for each station and each hour, we also have access to the original single trip data. This allows to fit our model based on the station feed data $C_{i,t}$ and compare our fit with the original data $\mathcal{N}_{ij,t}$. Hence, we can evaluate our model and check its performance to estimate the true trip counts $\mathcal{N}_{ij,t}$.

\subsection{Data Description}

\begin{figure}[t]
\center
\includegraphics[width = 0.49\textwidth]{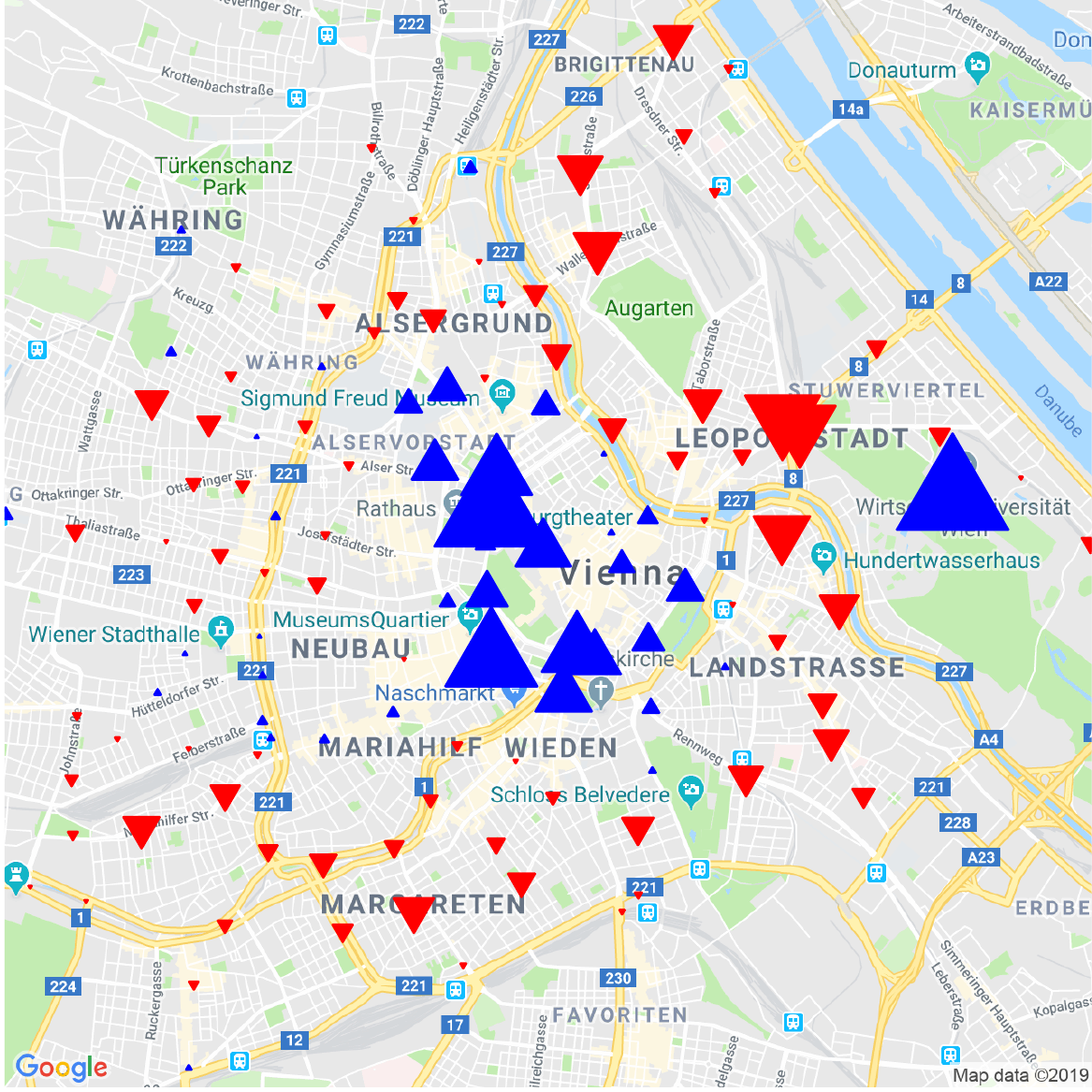}
\includegraphics[width = 0.49\textwidth]{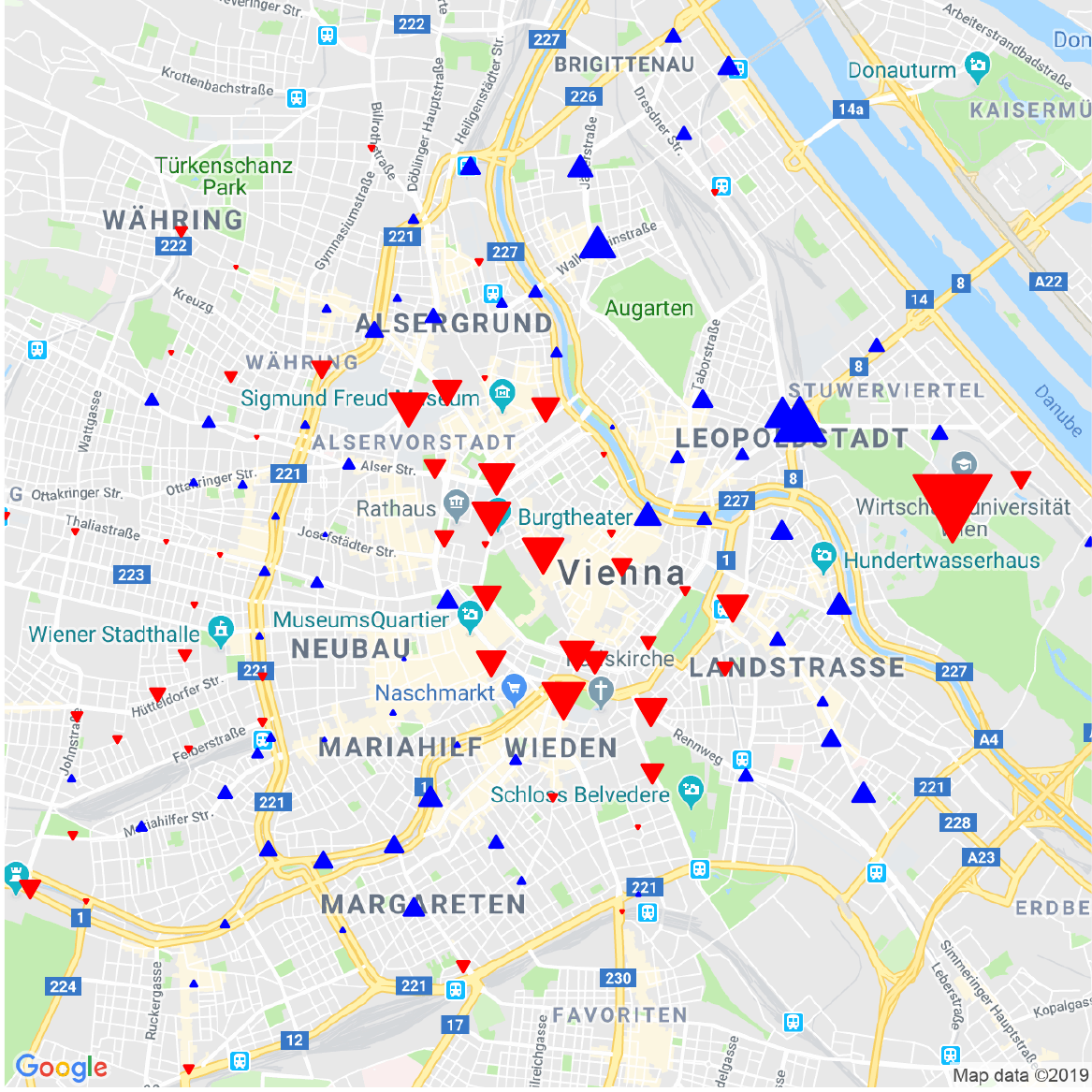}
\caption{Bike Stations in the Vienna Bike-Sharing System from 9-10 am (left panel) and from 5-6 pm (right panel). Blue (red reverse) triangles indicate stations that fill (clear) on average. The side lengths of the triangles are proportional to the absolute average differences of station feeds during that hour of the day. The plots are created using the \textbf{R}-package \texttt{ggmap} (\citeauthor{ggmap}, \citeyear{ggmap}).}
\label{fig: Vienna}
\end{figure}

In 2014, a total of $980 \text{ }360$ rides of customers was recorded. Additionally, the provider repositioned $79 \text{ }122$ bikes, e.g. to redistribute bikes from (almost) full to (almost) empty stations. These actions are denoted as service rides where most of them occurred in the morning hours. Since we are not able to distinguish between rides of customers and rides of the provider, our benchmark is always the sum of those. Only around 2/3 of the trips depart and end within the same hour. That is the reason why we installed the latent station $w$ which respects trips exceeding one time interval $[t-1, t)$. A major characteristic of bike-sharing networks is the sparsity. In our network, 99.1\% of the observed $\mathcal{N}_{ij,t}$ are equal to zero.

\begin{figure}[t]
\center
\includegraphics[width = 0.49\textwidth]{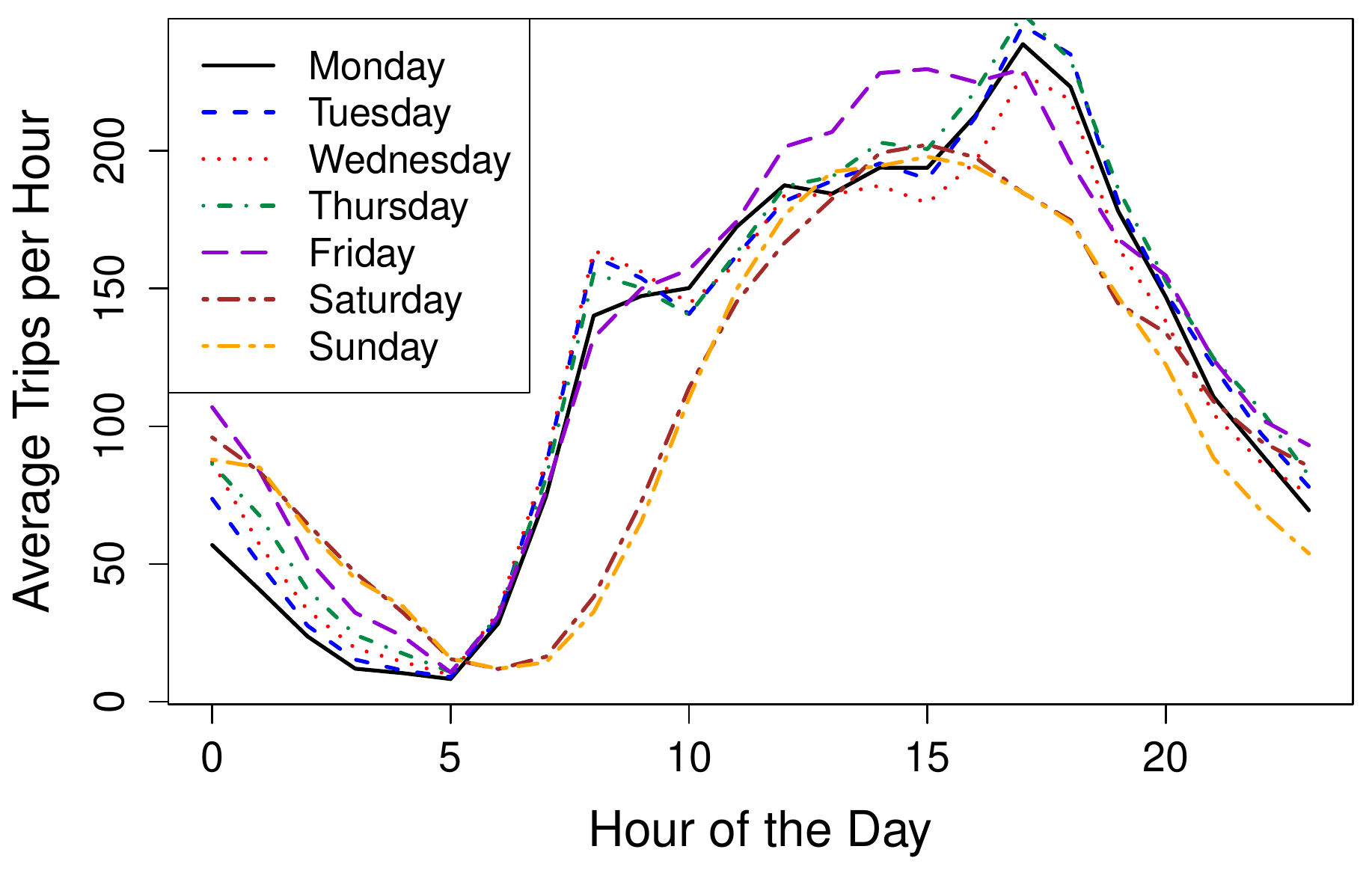}
\includegraphics[width = 0.49\textwidth]{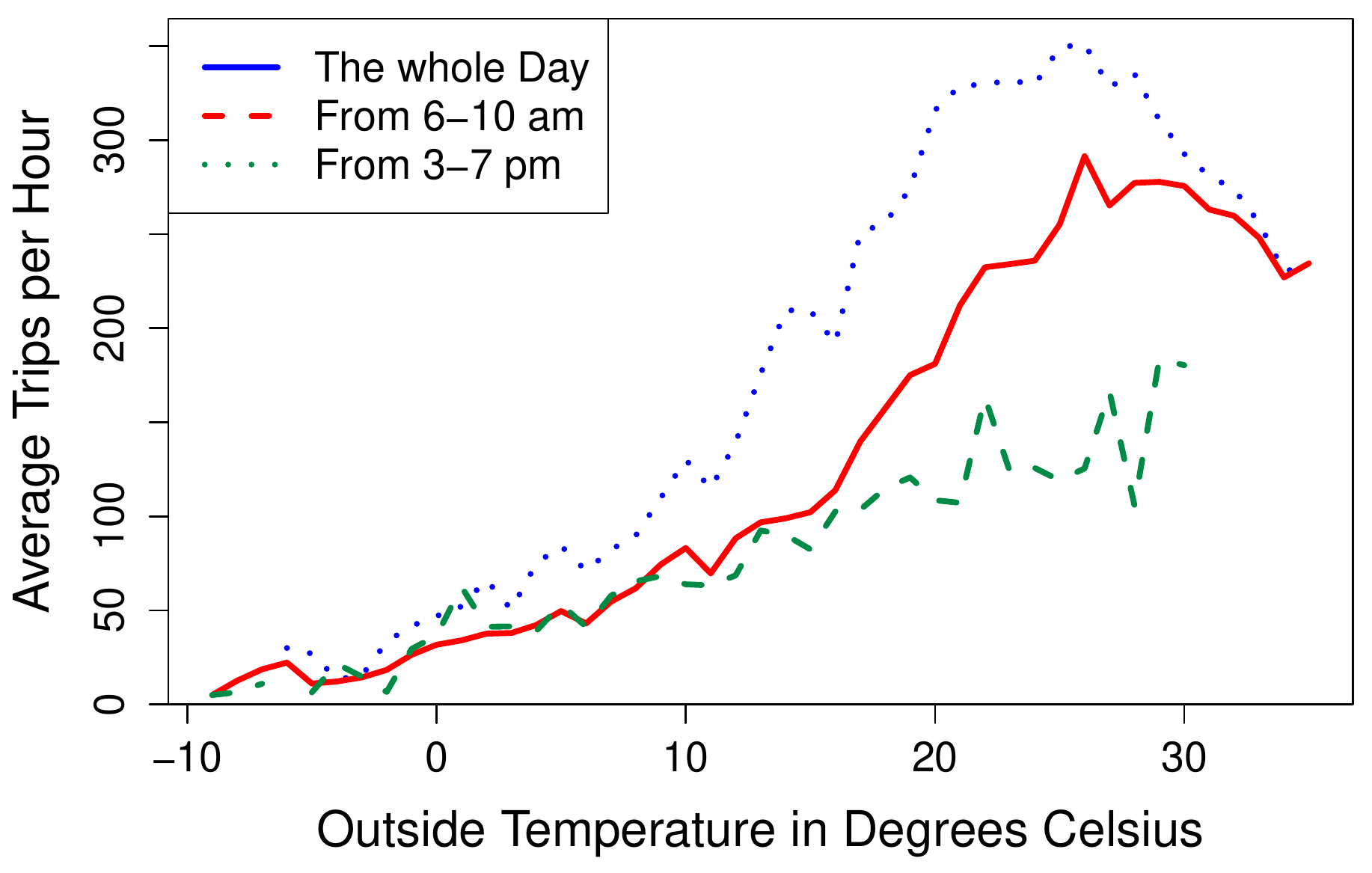}
\caption{Left Panel: Average Trips per Hour depending on Day of the Week and Time of the Day; Right Panel: Average Trips per Hour depending on the outside Temperature}
\label{fig: rides day temperature}
\end{figure} 

Figure \ref{fig: Vienna} gives an overview of the Vienna Bike-Sharing System. We can see that in the morning from 9 am to 10 am, stations in the city center and the station near the university have much more incoming than outgoing traffic. In the evening hours, we can observe the inverse effect and stations in the outskirts fill. The average absolute differences of the station feeds are larger between 9 and 10 am when compared to the hour from 5 to 6 pm. However, in the latter hour the Vienna Bike-Sharing Network is used more often which can also be observed from the left panel of Figure \ref{fig: rides day temperature} which shows that the average usage behavior depends heavily on the time of the day. Most of the trips occur in the evening hours and least in the early morning. Furthermore, we can see that particularly from 5 to 10 am the network is used less at the weekend than on weekdays. From previous studies on bicycle commuting (e.g. \citeauthor{smith2011bicycle}, \citeyear{smith2011bicycle}) we know that weather specific variables such as temperature or precipitation have a huge impact on the decision of people choosing the bike as mean of transportation or not. For the effect of the temperature, this is already illustrated in the right panel of Figure \ref{fig: rides day temperature}. 

\begin{figure}[t]
\center
\includegraphics[width = 0.49\textwidth]{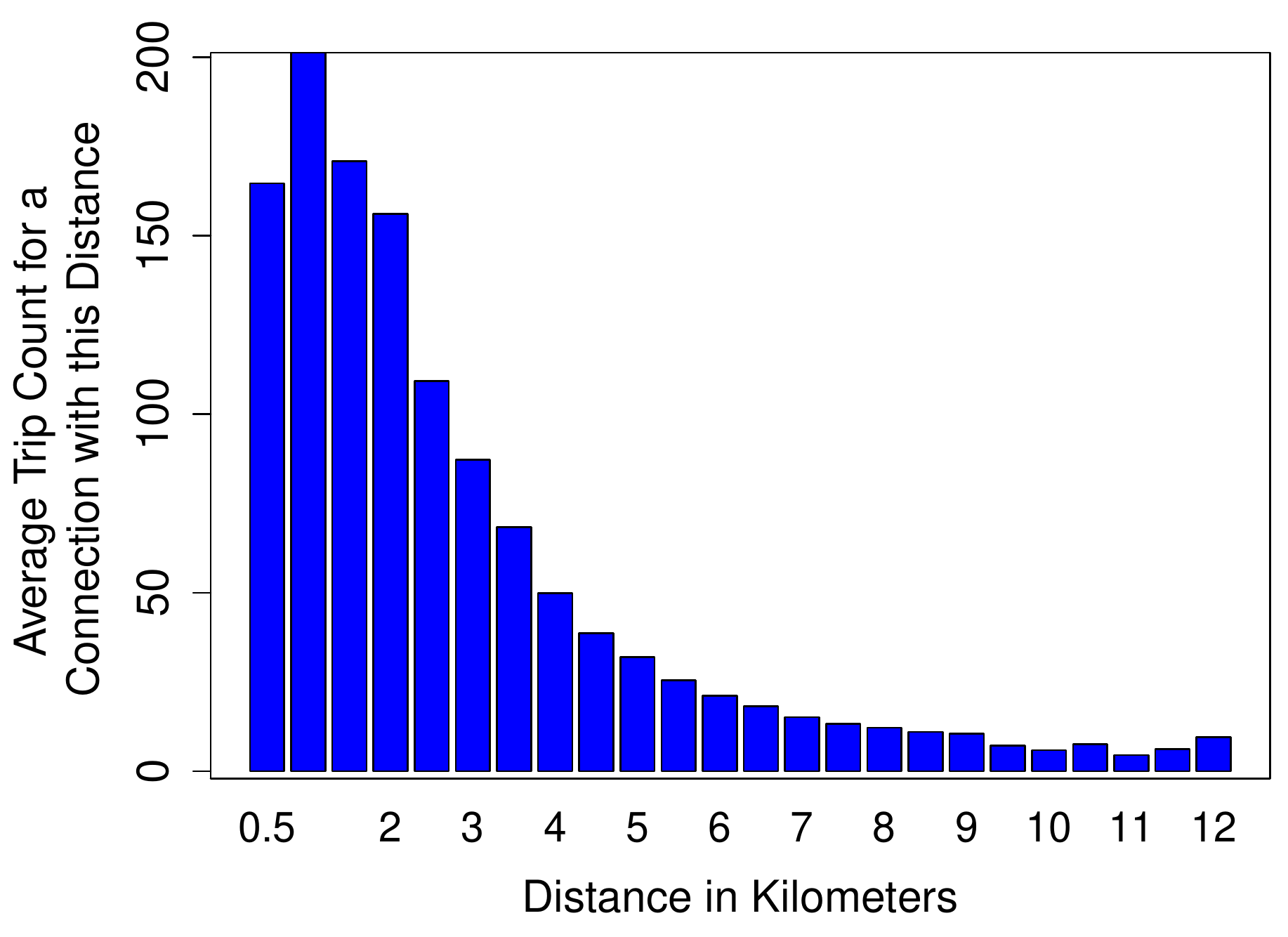}
\includegraphics[width = 0.49\textwidth]{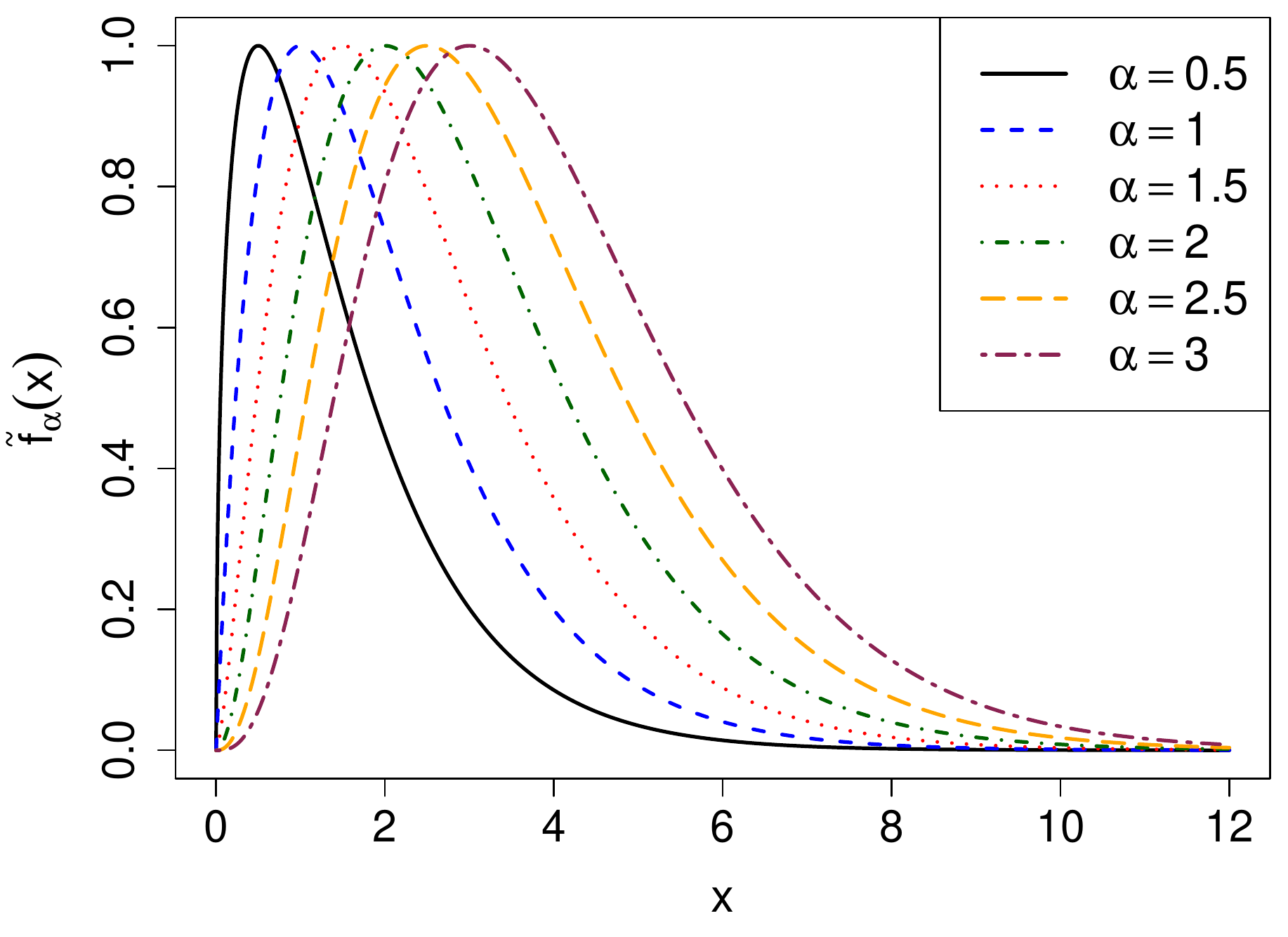}
\caption{Left Panel: Barplot of average trip counts dependent on the distance (aggregated to intervals of length 0.5 km) between two stations, loop counts are excluded. Right panel: Proposed non-linear effect $f_\alpha$ of the distance, functions are normalized to be one at the maximum.}
\label{fig: distance}
\end{figure}

The geo coordinates of the bike stations are used to compute the cycling distances $\texttt{dist}_{ij}$ between each of the station pairs $(i, j)$ as dyadic-specific covariates. The distance between station $i$ and station $j$ is defined as the shortest cycling route that Google Maps\footnote{\url{https://www.google.com/maps}} finds. This process can be automatized using an API key\footnote{\url{https://developers.google.com/maps/documentation/embed/get-api-key}}. Note that the set of distances $\lbrace \texttt{dist}_{ij} \rbrace$ does not define a metric, since in general $\texttt{dist}_{ij} \neq \texttt{dist}_{ji}$, e.g. due to one-way roads. As one could expect, the distance between two stations has a nonlinear effect on the trip counts when ignoring other effects, see the left panel of Figure \ref{fig: distance}. It turns out that spline-based modeling of dyadic covariates with our model leads to heavy inaccuracy of the estimates. To remedy this problem, we propose to transform the covariate $\texttt{dist}$ nonlinearly by $f_\alpha(\texttt{dist}) = \texttt{dist}^\alpha \exp (-\texttt{dist})$ for $\alpha > 0$ where the function $f_\alpha$ reaches its maximum at $\alpha$. Thus, this transformation respects the low count of trips for connections with a very short or a very high distance. In the right panel of Figure \ref{fig: distance} we plot the normalized functions $f_\alpha/\max(f_\alpha)$ for some values of $\alpha$. These functions behave accordingly to the diagram in the left plot of Figure \ref{fig: distance} which reflects the total trip counts in the year 2014 in dependence of the distance between tow stations.

\subsection{Model Implementation and Results}
\label{subsec: results}

We exemplarily estimate our model for the hour from 5-6 pm on weekdays.  This means that the model is estimated for $|\mathcal{T}| = 261$ days. In the chosen hour of the day, the network is least sparse but the average observed trip count is still only 0.017 and $98.5\%$ of the observed $n_{ij,t}$ are equal to zero. In the considered time frame, service rides of the provider account for 3\% of the trips such they should only slightly influence the estimates. We fit the dyadic model with parameters set to
\begin{align}
\begin{split}
\label{eq: param_1}
\mu_{\sbullet i, t} &= \sum_{j=1}^{N} \nu_{ji, t} + \nu_{wi, t}, \quad \mu_{i \sbullet , t} = \sum_{j=1}^{N} \nu_{ij, t} + \nu_{iw, t}, \\
\nu_{ij, t} &= \exp \left( \mathbold{z}_{ij, t}^\text{lin}\mathbold{\beta}^{\text{dyad}} + s_1^{\text{dyad}}(\texttt{temp}_t) + s_2^{\text{dyad}}(\texttt{seas}_t) + u_i^\text{out} + u_j^\text{in} \right).
\end{split}
\end{align}
Here, $\mathbold{z}_{ij, t}^\text{lin}$ represents the vector of covariates which are explicitly listed in the first column of Table \ref{tab: linear_effects}. In \eqref{eq: param_1}, $\texttt{temp}_t$ is the outside temperature in degrees Celsius at time $t$ and $\texttt{seas}_t$ is a value in the unit interval representing the time of the year. Note that weather covariates and calendar covariates are time specific, i.e. using the notation from above we collect these in $\mathbold{x}_t$. Outgoing and ingoing station specific covariates are listed as $\mathbold{x}_{i,t}$ and $\mathbold{x}_{j,t}$, respectively. The route specific covariate is dennoted as $x_{ij}$, which in this case does not depend on $t$. We fit two models, one with the route specific covariate included, with estimates denoted as $\widehat{\mathbold{\beta}}^\text{dyad}$, and one by omitting the route specific quantities which simplifies the model to the station specific type explained in Section \ref{subsec: station based model}. Estimates in this model are denoted as $\widehat{\mathbold{\beta}}^\text{station}$.

\begin{table}[t]
\centering
\caption{\label{tab: linear_effects} Estimates of fixed linear effects, standard errors in brackets.}
\resizebox{\textwidth}{!}{\begin{tabular}{rllrr|r}
\toprule
& Variable  & Explanation    & $\widehat{\mathbold{\beta}}^{\text{station}}$ & $\widehat{\mathbold{\beta}}^{\text{dyad}}$   & $\widehat{\mathbold{\beta}}^\text{Poisson}$ \\
\midrule	
& 1 & Intercept                                         & 0.789 (0.176)  & -4.896 (0.473)   & \sl{-4.607 (0.249)} \\ 
\ldelim\{{7}{3mm}[$\mathbold{x}_t$] & $\texttt{rain}$ &  Relative duration in $[t-2,t)$   & -1.041 (0.045) & -1.042 (0.045)   & \sl{-1.083 (0.030)} \\
& $\texttt{sun}$ & Relative duration in $[t-2,t)$     & 0.349 (0.038)  &  0.349 (0.034)   & \sl{0.193 (0.015)} \\
& $\texttt{tue}$ & Tuesday                            & 0.058 (0.027)  &  0.058 (0.027)   & \sl{0.048 (0.013)} \\
& $\texttt{wed}$ & Wednesday                          & 0.029 (0.027)  &  0.029 (0.027)   & \sl{-0.021 (0.013)} \\
& $\texttt{thu}$ & Thursday                           & -0.033 (0.026) & -0.037 (0.026)   & \sl{0.001 (0.013)} \\
& $\texttt{fri}$ & Friday                             & -0.010 (0.026) & -0.013 (0.026)   & \sl{-0.103 (0.013)} \\
& $\texttt{ph}$ & Public Holiday                      & -0.136 (0.041) & -0.126 (0.041)   & \sl{-0.187 (0.024)}   \\
\ldelim\{{2}{5mm}[$\mathbold{x}_{i,t}$] & $\texttt{nobikes}$ & No bikes in $t-1$ and $t$  & -0.948 (0.066) & -0.840 (0.063)   & \sl{-1.181 (0.038)} \\
& $\texttt{hubout}$ & Log-distance from next hub    & -0.340 (0.088) & -0.357 (0.088)   & \sl{-0.266 (0.062)} \\
\ldelim\{{2}{5mm}[$\mathbold{x}_{j,t}$] & $\texttt{noboxes}$ & No boxes in $t-1$ and $t$  & -0.668 (0.061) & -0.575 (0.060)   & \sl{-0.936 (0.040)} \\
& $\texttt{hubin}$ & Log-distance to next hub       & -0.353 (0.089) & -0.343 (0.090)   & \sl{-0.258 (0.067)} \\
\ldelim\{{1}{4mm}[$x_{ij}$]  & $f_{1.7}(\texttt{dist})$ & Non-linear transformation of distance  
                                                      &                & 3.872  (1.417)   & \sl{3.681 (0.032)} \\
\bottomrule
\end{tabular}}
\end{table}

In Table \ref{tab: linear_effects} we summarize the dyadic model's estimates  $\widehat{\mathbold{\beta}}^{\text{dyad}}$  and the station based model's estimates $\widehat{\mathbold{\beta}}^{\text{station}}$. For comparison, we also list the estimates $\widehat{\mathbold{\beta}}^\text{Poisson}$, if we fit the model to the original data $\mathcal{N}_{ij,t}$ with $\mathcal{N}_{ij,t} \sim \text{Poi}(\mu_{ij,t})$ using the original model \eqref{eq: pois_model} and taking the same covariates into account as in \eqref{eq: param_1}. These parameters may serve as benchmark since they rely on complete trip data. Remember that the aim of this paper is to estimate the model based on station feed data only.

The covariates $\texttt{rain}_t$ and $\texttt{sun}_t$ quantify the relative duration of rain and sunshine in the interval $[t-2, t)$, i.e. they take values in the interval $[0,1]$. The impact of both is clearly significant, showing a negative effect for $\texttt{rain}_t$ and a positive effect for $\texttt{sun}_t$. For the weekday effect, we use dummy coding. The results show that the use of the bike-sharing system varies slightly during the week with Tuesday being the most frequented day conditional on all other effects. However, on public holidays, indicated by $\texttt{ph}_t$, there are significantly less trips. The time- and station dependent covariates $\texttt{nobikes}_{i,t}$ and $\texttt{noboxes}_{j,t}$ indicate whether there have been no bikes or boxes available at time points $t-1$ and $t$, i.e. at the beginning and the end of a time interval. As one could expect, the corresponding parameters are clearly significant with a negative sign. The covariates $\texttt{hubout}_{i,t}$ and $\texttt{hubin}_{j,t}$ specify the logarithm of the distance (in 50 meters) of a station to the next underground or train station. Hence, the further away a bike station is from major public transportation hubs, the fewer the bike station is used. 

As motivated in the previous subsection (see also Figure \ref{fig: distance}), we respect the nonlinear effect of the dyadic specific variable $\texttt{dist}_{ij}$ by making use of the transformation $f_\alpha(\texttt{dist}_{ij})$. In order to find a proper value for $\alpha$, we first included several basis functions $f_{\alpha_1},\dots,f_{\alpha_K}$ into the linear predictor with zero as a lower bound for the respective parameter estimates. However, most of the estimates were at the boundaries with high standard deviations. Hence, we fitted the model with different sets of basis functions leading to $\alpha = 1.7$ as a reasonable value, i.e. we estimate trip lengths of around 1.7 kilometers to be most likely. This is in accordance with the terminology ``first/last mile problem'' (e.g. \citeauthor{shaheen2010bikesharing}, \citeyear{shaheen2010bikesharing}). The nonlinear effect of the distance between two stations that was estimated with the dyadic model is not significantly different from the effect that was estimated with the Poisson model. This is also illustrated in the left panel of Figure \ref{fig: dyad model} in Appendix \ref{app: model evaluation}.

\begin{figure}[t]
\centering
\includegraphics[width = 0.49\textwidth]{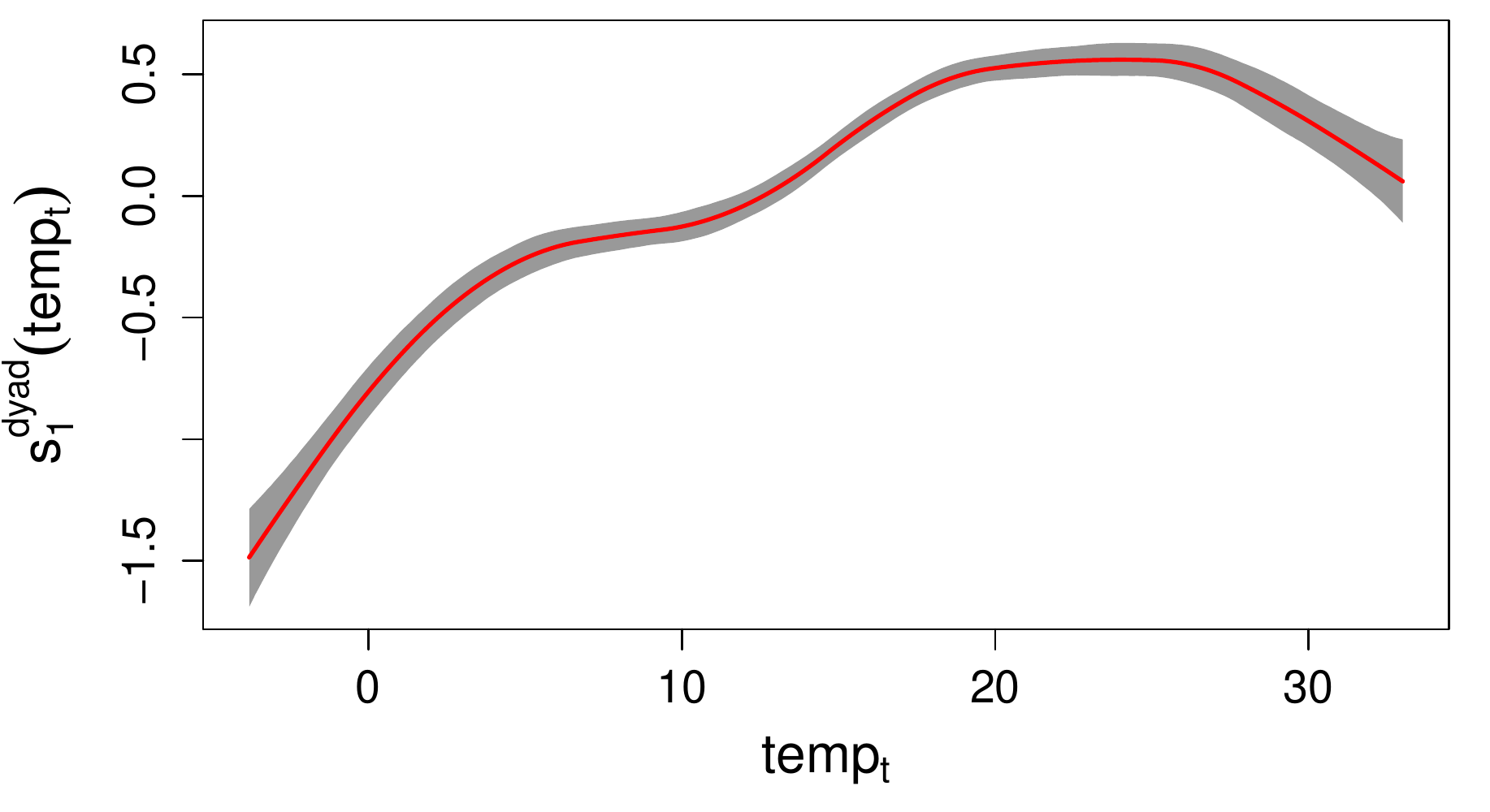}
\includegraphics[width = 0.49\textwidth]{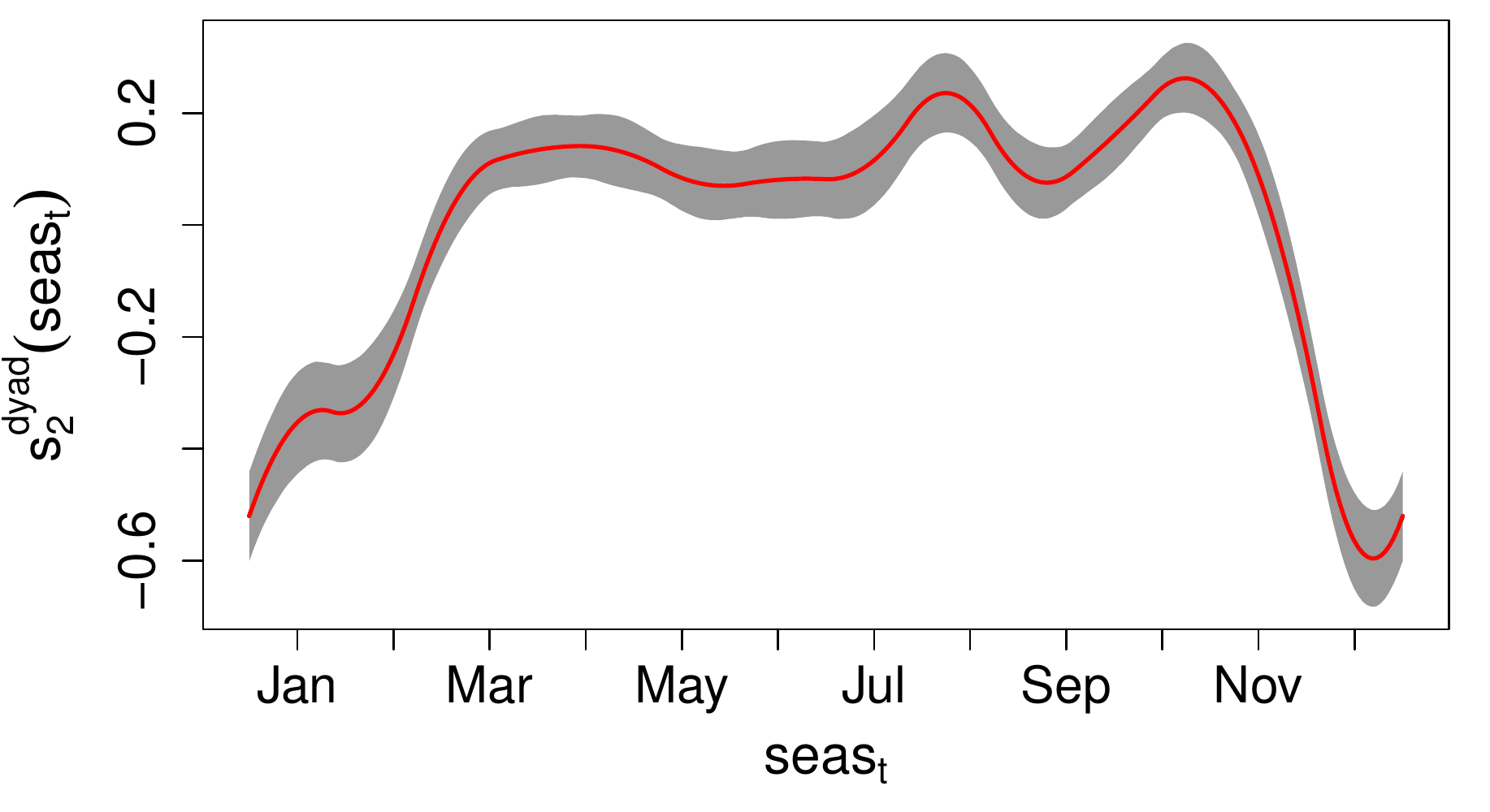}
\includegraphics[width = 0.49\textwidth]{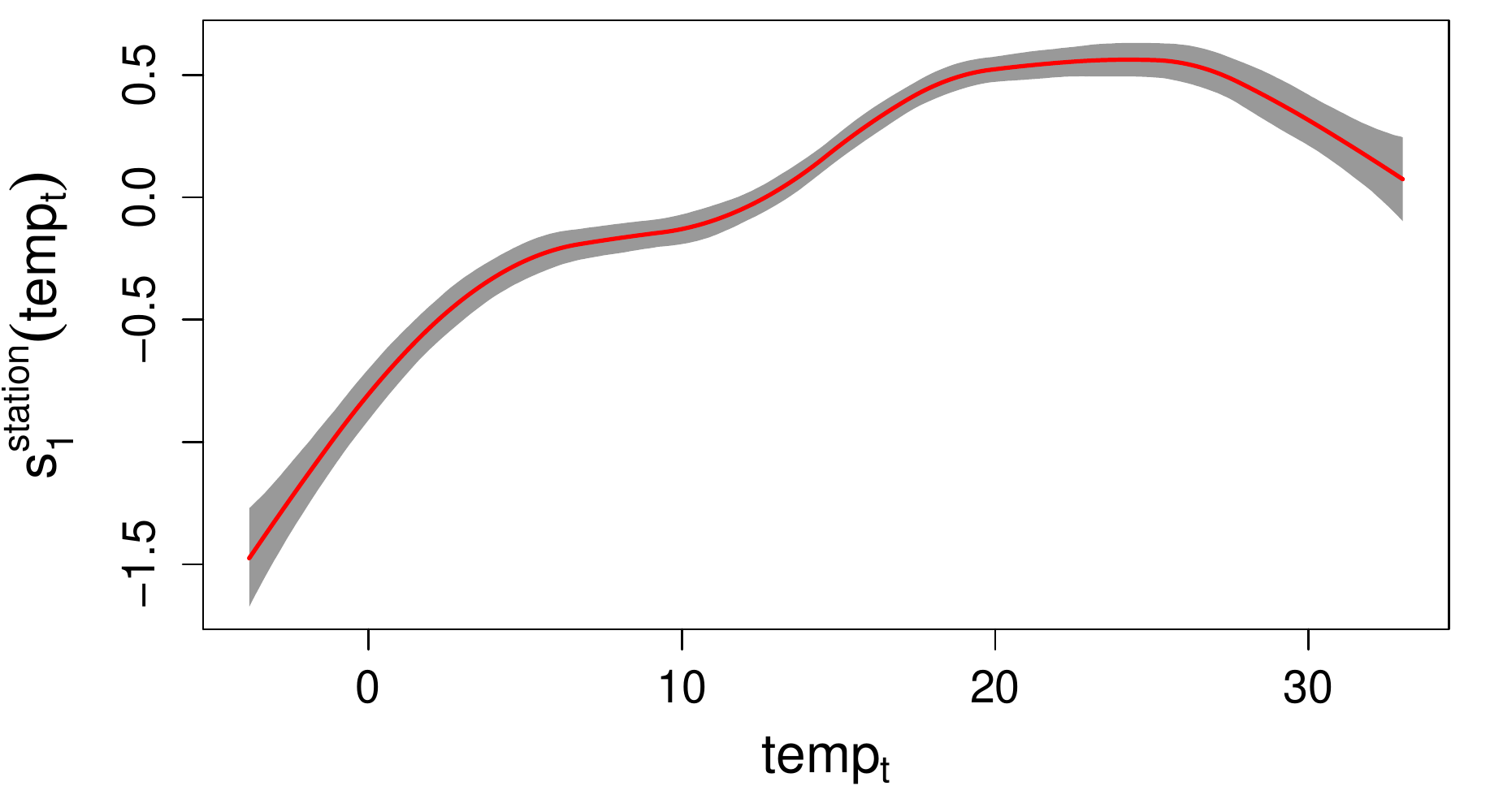}
\includegraphics[width = 0.49\textwidth]{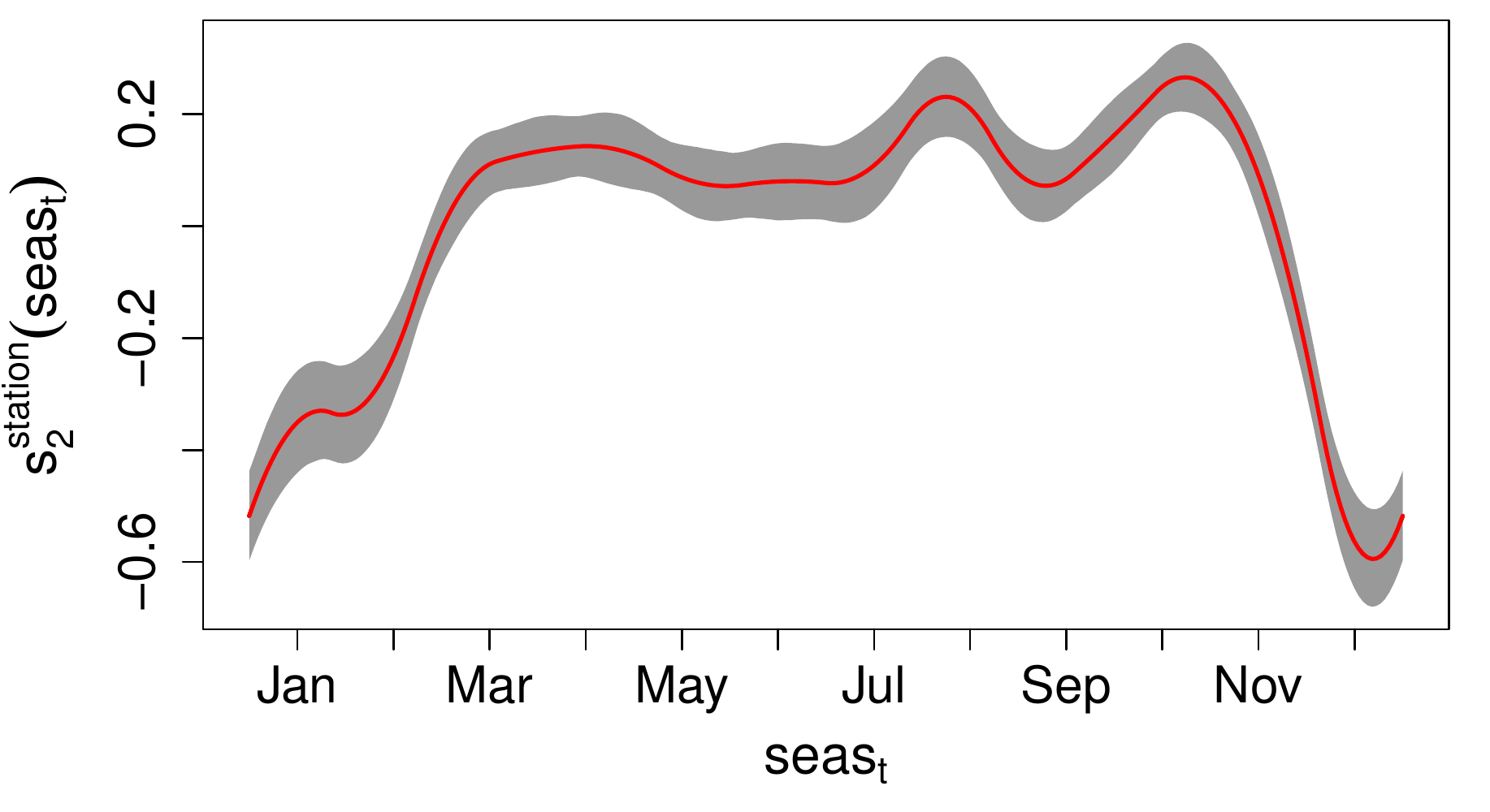}
\caption{\label{fig: smooth_effects} Estimates of smooth functions with 95\% Confidence Bands. The top row represents the dyadic model and the bottom row represents the station based model.}
\end{figure}

The fitted smooth functions for temperature and seasonality are shown in Figure \ref{fig: smooth_effects} including 95\% confidence bands. In order to calculate the confidence bands, we simulated 10 000 times from the distribution of the estimated spline parameters $\widehat{\mathbold{\gamma}}_m$. The lower and upper confidence bands are hence determined by the respective pointwise 2.5\% and 97.5\% quantiles. For both parameterizations, i.e. station based and dyadic based, the smooth functions $s_1(\texttt{temp}_t)$ and $s_2(\texttt{seas}_t)$ look similar. As one could expect from Figure \ref{fig: rides day temperature}, the expected count of trips increases with the temperature, but if it is too hot outside, the effect reverses. The confidence bands are wider for very low and very large temperatures due to fewer observations. The right panels show that the system is mostly used between April and October, disregarding all other effects included in the model. However, the estimated effect is much more wiggly than the temperature effect.

\begin{figure}[t]
\centering
\includegraphics[width = 0.49\textwidth]{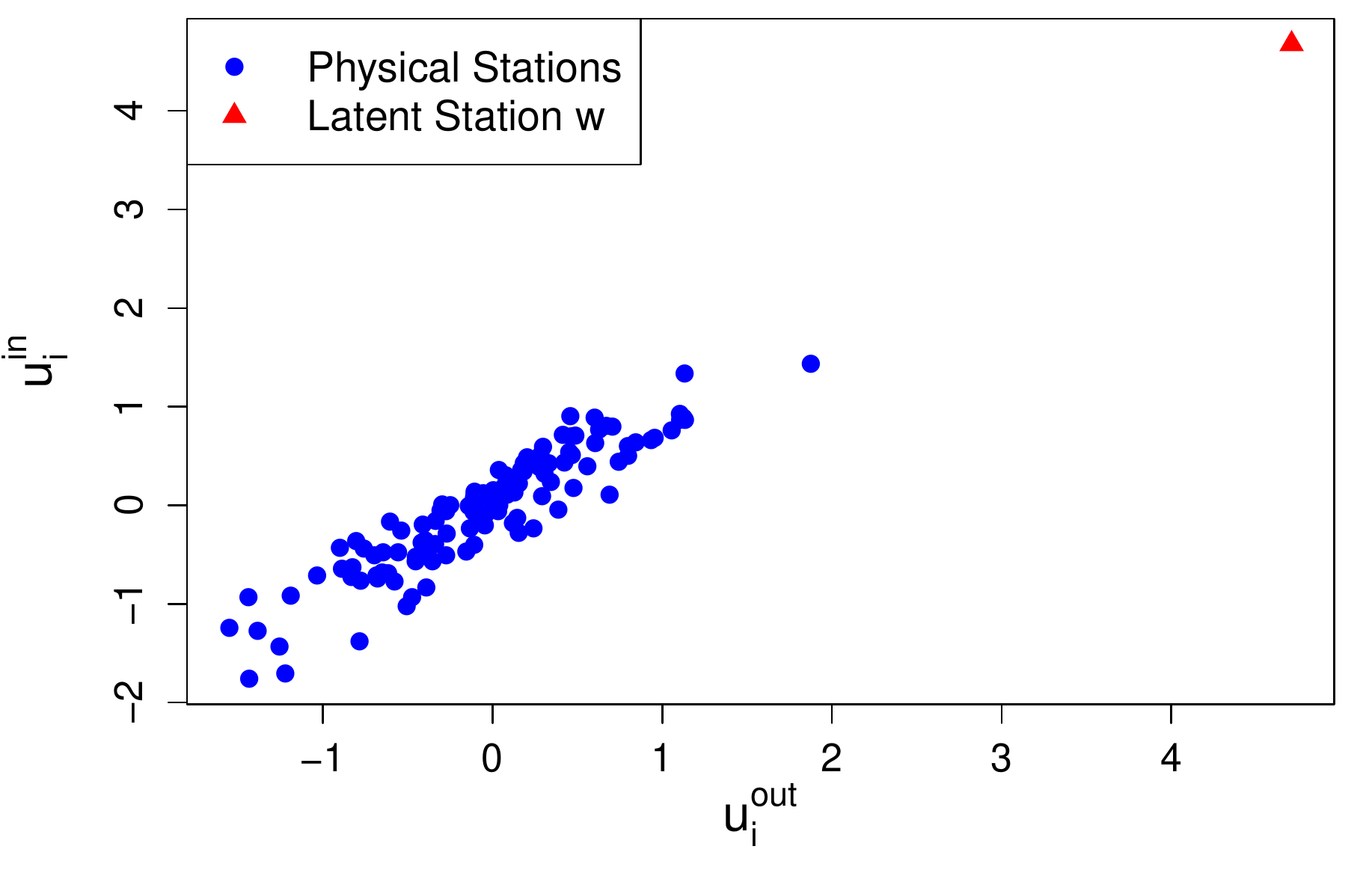}
\includegraphics[width = 0.49\textwidth]{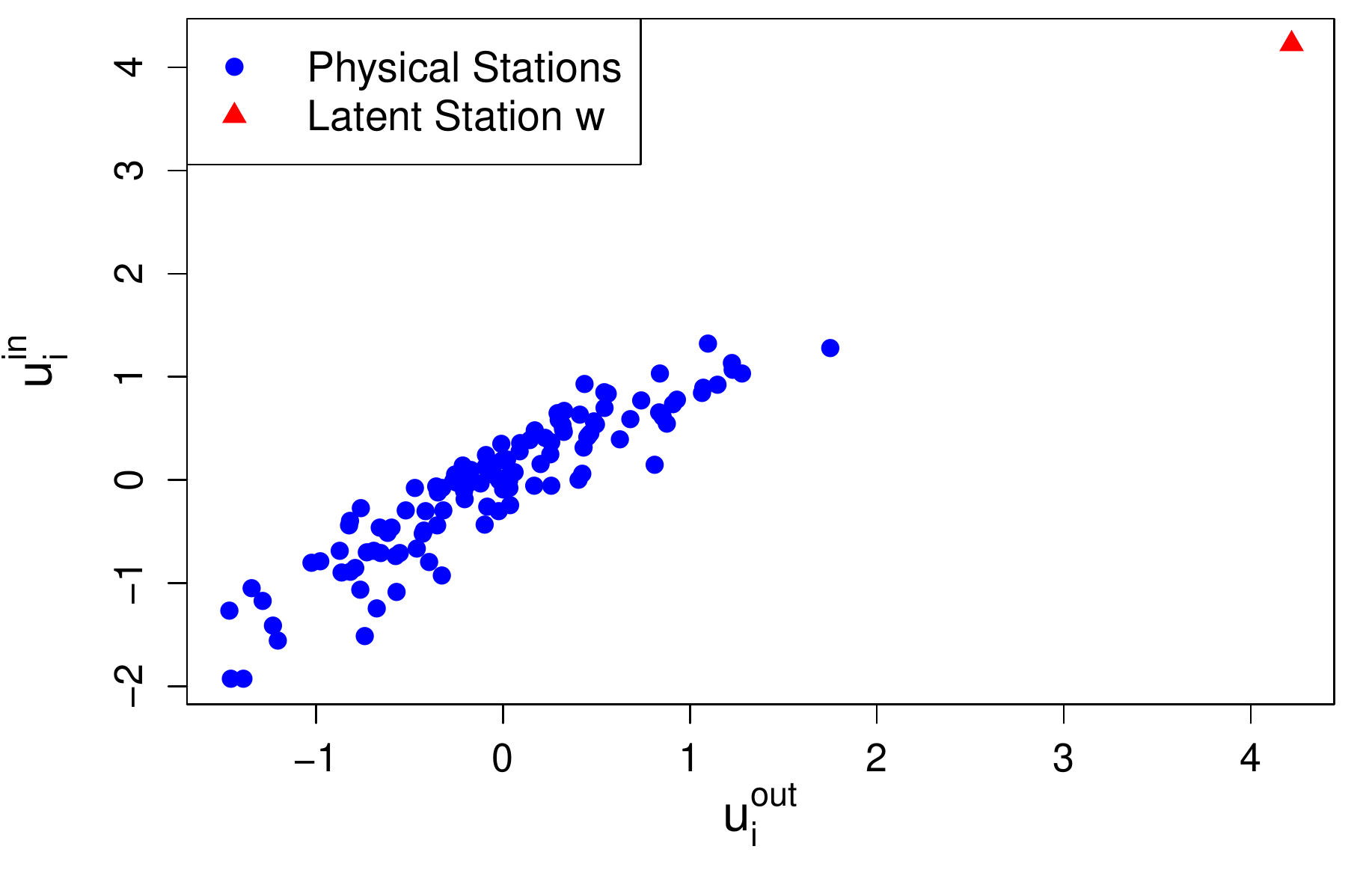}
\caption{\label{fig: random_effects} Estimates of Random Effects with the Dyadic Model (Left Panel) and the Station Based Model (Right Panel)}
\end{figure}

In Figure \ref{fig: random_effects} we depict the estimated random effects for both the dyadic model and the station based model. The estimates of $\mathbold{u}_i$ for $i = 1,\dots,N$  fit to the specified distribution \eqref{eq: distribution_random_effects}. The isolated vector $\mathbold{u}_w$ in the upper right corner of both plots reflects the large count of trips which either started in a previous time interval or reached their destination not before the subsequent time interval. The estimated covariance matrices $\widehat{\mathbf{\Sigma}}^{\text{dyad}}$ and $\widehat{\mathbf{\Sigma}}^{\text{station}}$  which are given by
\begin{equation*}
\widehat{\mathbf{\Sigma}}^{\text{dyad}} = \begin{pmatrix}
 0.604 & 0.560 \\
0.560 & 0.585
\end{pmatrix}, \quad  \widehat{\mathbf{\Sigma}}^{\text{station}} = \begin{pmatrix}
 0.590 & 0.573 \\
 0.573 & 0.631
\end{pmatrix}
\end{equation*}
respectively, are very similar. The components of the random vectors $\mathbold{u}_i$ are highly correlated which can also be observed in Figure \ref{fig: random_effects}. Hence, stations with many incoming trips tend to have also a clearly high outgoing trip count. The variances are quite similar, and thus the distributions of $u^{\text{in}}$ and $u^{\text{out}}$ are similar, too.

Finally, we do compare the estimates $\widehat{\mathbold{\beta}}^{\text{dyad}}$ and $\widehat{\mathbold{\beta}}^{\text{Poisson}}$, where the latter are fit directly to the trips and hence serve as benchmark, if original trips counts instead of station feeds are available. We see in Table \ref{tab: linear_effects} a general concordance though some parameter estimates differ in size. Major differences occur for the sunshine variable and the indicator whether a station ran out of bikes or is full. The difference for the latter two variables is not surprising since these are very dynamic covariates, i.e. a station can be empty now and offer bikes again a couple of minutes later. Apparently, the standard errors are larger for the Skellam models compared to the full trip data fit. The estimates and standard errors that were fit to the Skellam models are very similar except for the intercept. However, this can can be explained by the derivation of the station based model in Section \ref{subsec: station based model}. 

\subsection{Model Evaluation}

\begin{figure}[t]
\centering
\includegraphics[width = \textwidth]{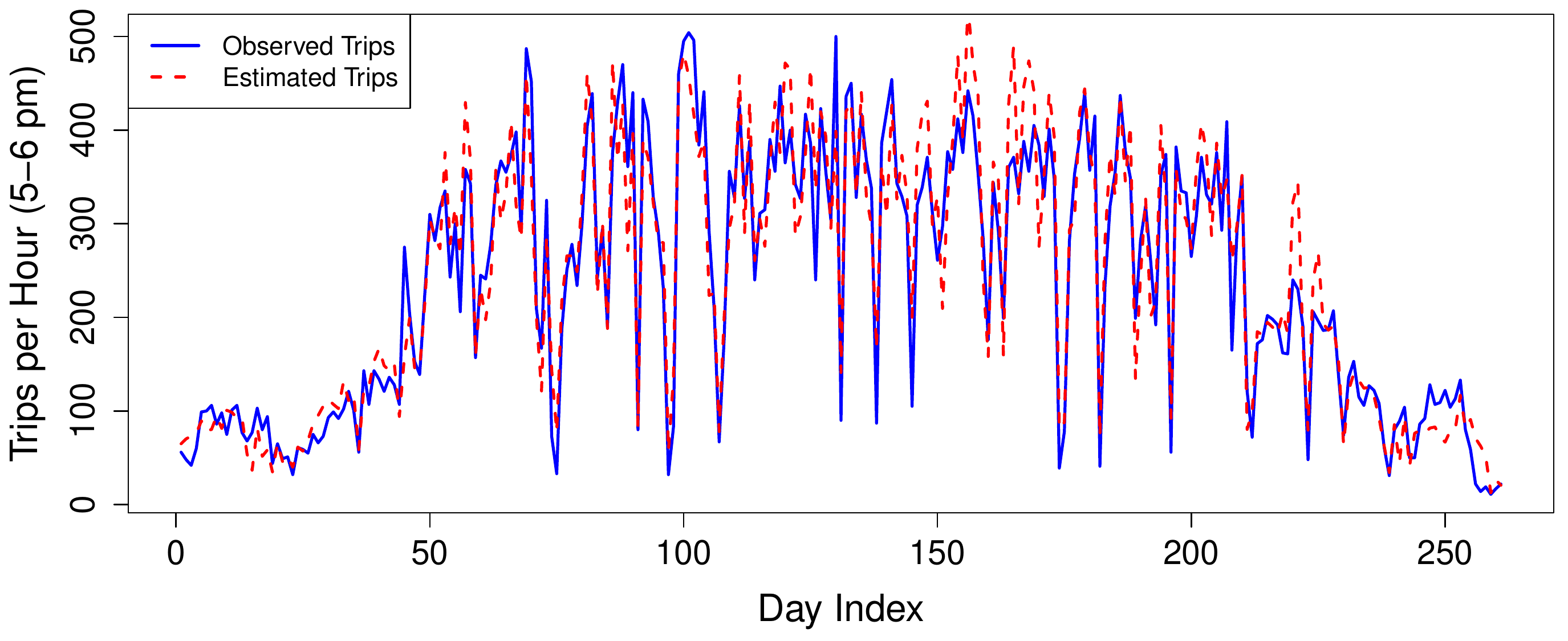}
\includegraphics[width = 0.49\textwidth]{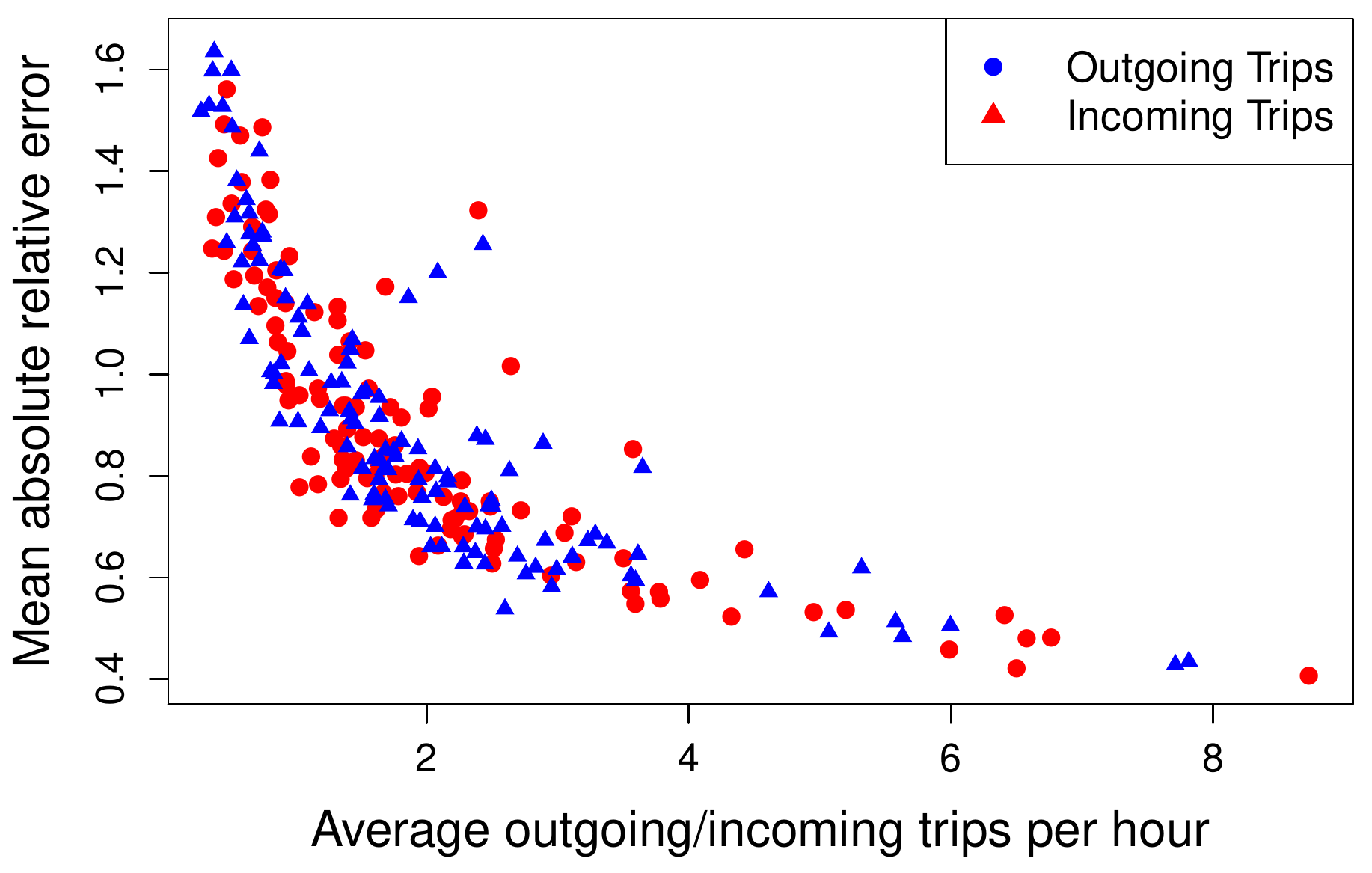}
\includegraphics[width = 0.49\textwidth]{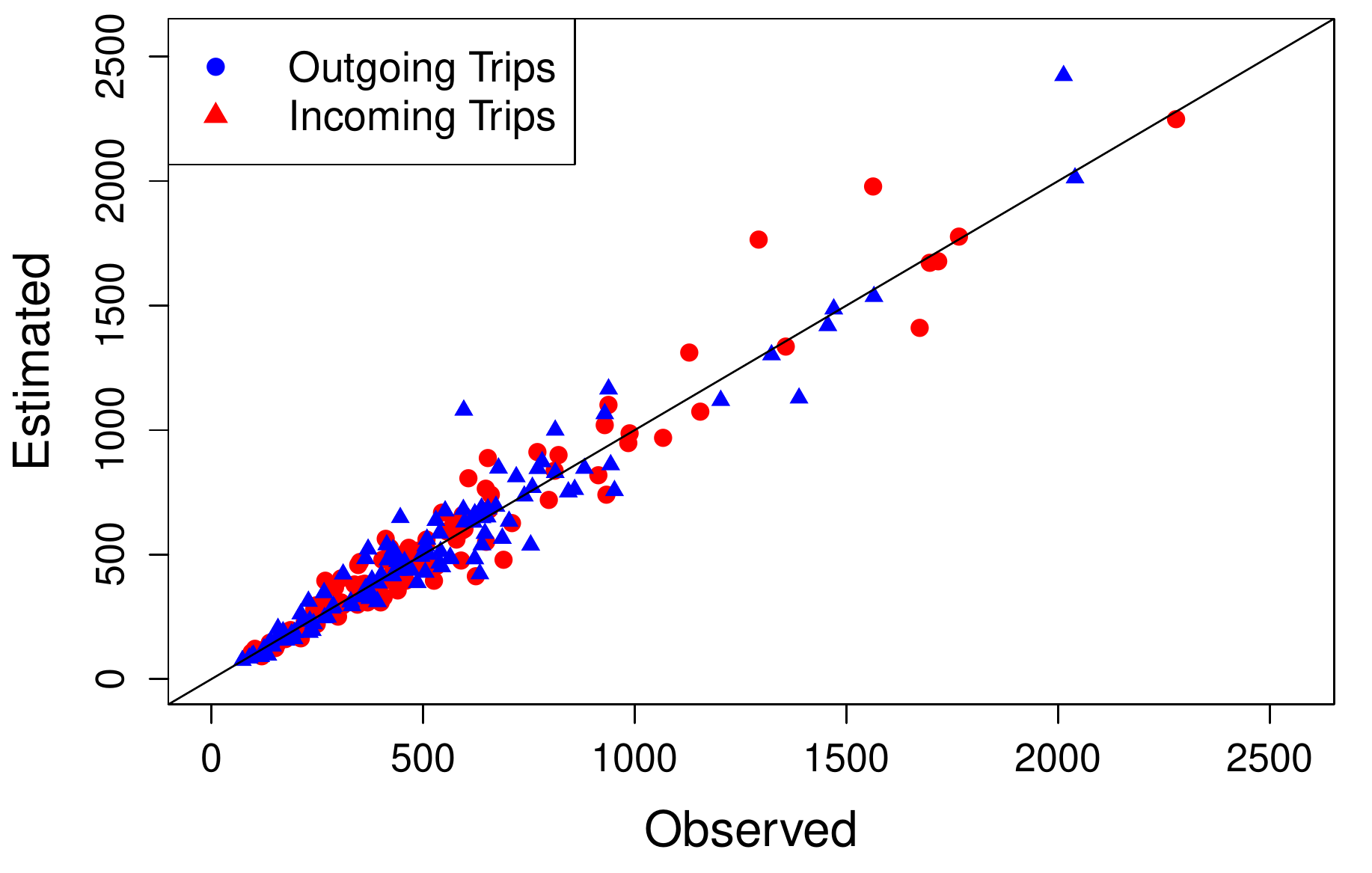}
\caption{\label{fig: trips_per_day_station} Top Panel: Observed vs. Estimated Trip Count per Hour; Bottom Left Panel: Means of relative Errors $\Delta_{i \sbullet, t}$ and $\Delta_{ \sbullet i, t}$ for every Station $i$; Bottom Right Panel: Observed vs. Estimated Cumulated In- and Outdegrees.}
\end{figure}

The above comparison of the Skellam model with the Poisson model fit to the separate trip data serves already as model evaluation and confirms that station feeds allow to obtain information about unobserved network flows. We will now further investigate the performance of the model. To do so we first look at the ability of predicting in- and outdegrees with the station based model. In the top panel of Figure \ref{fig: trips_per_day_station} we depict the total count of outgoing trips in the chosen hour of the day depending on the day index. With just few exceptions, the relative errors
\begin{equation*}
\Delta_t^\text{out} = \left| \sum_{i=1}^N \widehat{\mu}_{i \sbullet, t} - \sum_{i=1}^N \mathcal{N}_{i \sbullet, t} \right| / \sum_{i=1}^N \widehat{\mu}_{i \sbullet, t}, \quad t \in \mathcal{T}
\end{equation*}
are rather small with a mean absolute relative error of 0.169. The corresponding relative errors $\Delta_t^\text{in}$ have a mean of 0.175. 

Next, we consider the relative station-wise  errors with respect to outgoing or incoming trips to or from station $i$, i.e.
\begin{equation*}
\Delta_{i \sbullet, t} = \left|  \widehat{\mu}_{i \sbullet, t} -  \mathcal{N}_{i \sbullet, t} \right| /  \widehat{\mu}_{i \sbullet, t}, \quad 
\Delta_{ \sbullet i, t} = \left|  \widehat{\mu}_{ \sbullet i, t} -  \mathcal{N}_{ \sbullet i, t} \right| /  \widehat{\mu}_{ \sbullet i, t}, \quad i = 1,\dots,N, \quad t \in \mathcal{T}.
\end{equation*}

In the bottom left panel of Figure \ref{fig: trips_per_day_station} we show the mean absolute relative errors of out- and indegrees for every station. Here, we can see that the more frequented a station is used, the lower is the prediction error. This is not surprising since the lower $\widehat{\mu}_{i\sbullet,t}$ or $\widehat{\mu}_{\sbullet i,t}$, the larger the relative effect of the absolute prediction error. It therefore appears more plausible to consider the estimates of the cumulated in- and outdegrees for every station $i$, that are 
\begin{equation*}
\widehat{\mu}_{i \sbullet, \sbullet} = \sum_{t \in \mathcal{T}} \widehat{\mu}_{i \sbullet, t}, \quad \widehat{\mu}_{ \sbullet i, \sbullet} = \sum_{t \in \mathcal{T}} \widehat{\mu}_{ \sbullet i, t}.
\end{equation*}
This is shown in the bottom right plot of Figure \ref{fig: trips_per_day_station} and there we see a promising concordance with the corresponding observed counts. The prediction errors are symmetrically around zero, i.e. there is no systematical bias.

\begin{figure}[t]
\centering
\includegraphics[width = 0.6\textwidth]{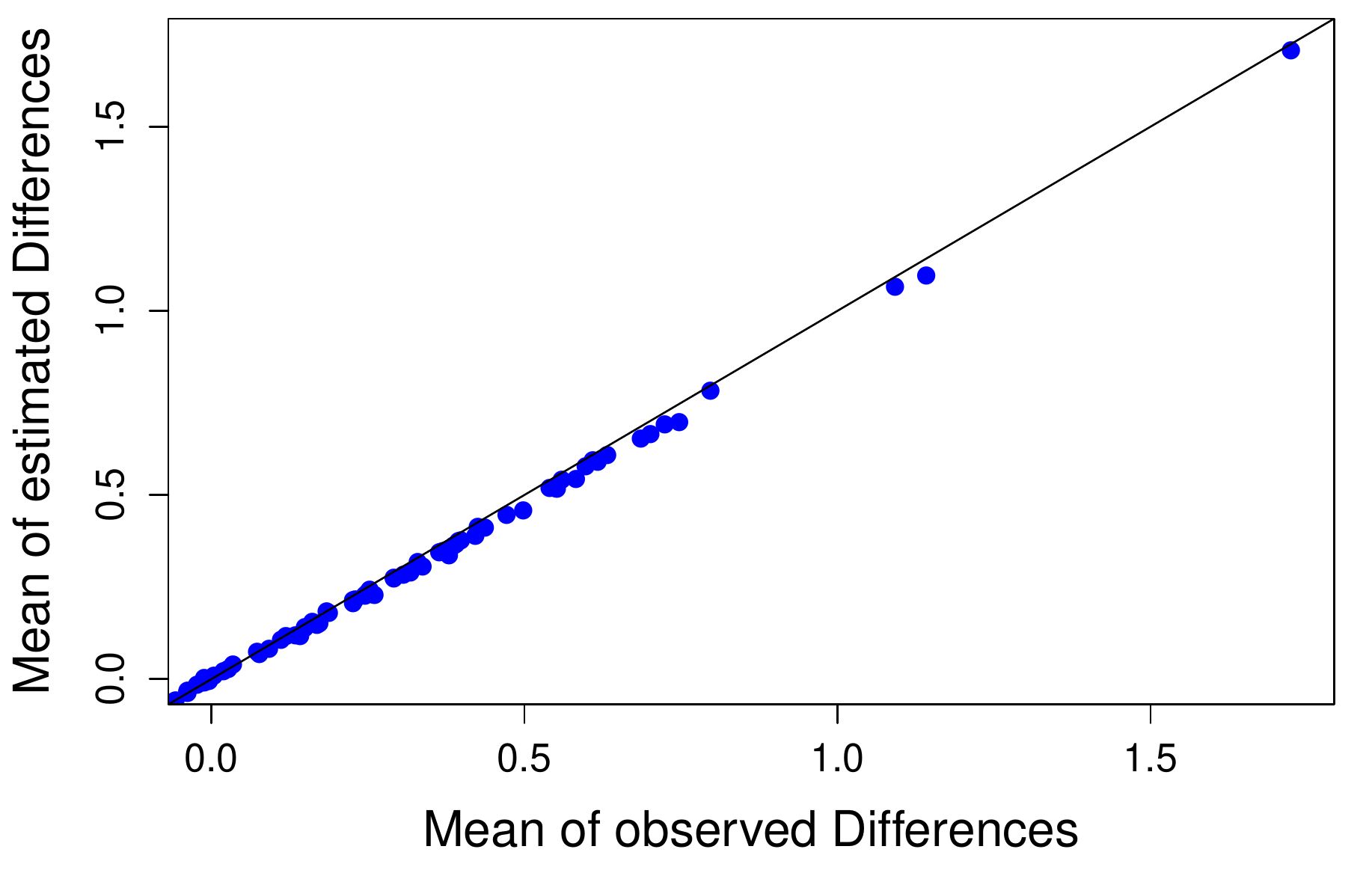}
\caption{\label{fig: differences} Observed vs. estimated Differences of Stations Feeds for every Station averaged over the observation Period}
\end{figure}

Since we actually fitted differences of station feeds we further compare  the actual differences $\mathcal{D}_{i,t}$ and the estimated differences $\widehat{\mathcal{D}}_{i,t} = \widehat{\mu}_{\sbullet i,t} - \widehat{\mu}_{i \sbullet, t}$. In the right panel of Figure \ref{fig: differences} we depict those for every station $i$ averaged over the observation period. The averaged observed differences are very close to the averaged estimated differences. Further evaluation of the dyadic Skellam model can be found in Appendix \ref{app: model evaluation}.

\section{Simulation Study}
\label{sec: simulation}

After having applied the model to real data, we evaluate its performance making use of simulated data, i.e. we compare the estimates $\widehat{\mu}_{ij,t}$ with the actual expected trip counts $\mu_{ij,t}$. We set up the following simulation scenario. For each simulated network we consider one fixed hour of the day on $| \mathcal{T}| = 500$ days and a trip  arrives its destination in the subsequent interval with probability $1/3$. Hence, we also include the latent station $w$ in the model which considers trips that exceed a single time interval. The count of bike stations $N$ is set to 20 and for simplicity we assume that the they can neither be empty nor full at any time.

The network flow is assumed to be dependent on two covariates which are included as linear effects in $\eta$ -- a time-dependent covariate  $z_t^{(1)}$ and a dyadic covariate $z_{ij}^{(2)}$. Additionally, an intercept and the two-dimensional node-dependent random effects $\mathbold{u}_i = (u_i^\text{out}, u_i^\text{in})^\top$ are included. The exogenous data are simulated independently according to $\mathbold{z}_{ij,t} = (z_t^{(1)}, z_{ij}^{(2)}) \sim \mathcal{N}(\mathbf{0}, \mathbold{I}_2)$ where we set $z_{ij}^{(2)} = z_{ji}^{(2)}$ for all $i,j = 1,\dots,N$ for simplicity. The random effects $\mathbold{u}_i = (u_i^\text{out}, u_i^\text{in})^\top$ are drawn independently from a bivariate normal distribution with mean $\mathbold{\mu} = (0,0)^\top$ and covariance matrix 
\begin{equation}
\label{eq: Sigma_simulation}
\mathbf{\Sigma} = \begin{pmatrix}
\sigma_1^2 & \sigma_{12}^2 \\ \sigma_{12}^2 & \sigma_2^2 
\end{pmatrix} = \begin{pmatrix}
1 & 0.9 \\ 0.9 & 1 
\end{pmatrix}.
\end{equation}
such that the expected counts of incoming and outgoing trips are clearly positively correlated. In the $s$-th simulation we generate the trip counts by $Y_{s,ij,t} \sim \text{Poi}(\mu_{ij,t})$ for $i,j \in 1,\dots,N$ and $t \in \mathcal{T}$ where
\begin{equation*}
\mu_{ij,t} = \exp \left( \beta_0 + \beta_1 z_t^{(1)} + \beta_2 z_{ij}^{(2)}  + u_i^{\text{out}} + u_j^\text{in} \right)  .
\end{equation*} 
Since incoming bike trips in $[t-1,t)$ might have departed in the previous time interval, we additionally need to simulate the trip counts of the preceding hour, denoted by $Y_{s,ij,t}^\star$. Here, we set $\mu_{ij,t}^\star = 0.9 \mu_{ij, t}$ which reflects clearing stations by trend during the evaluated hour of the day.

\begin{figure}[t]
\includegraphics[width = \textwidth]{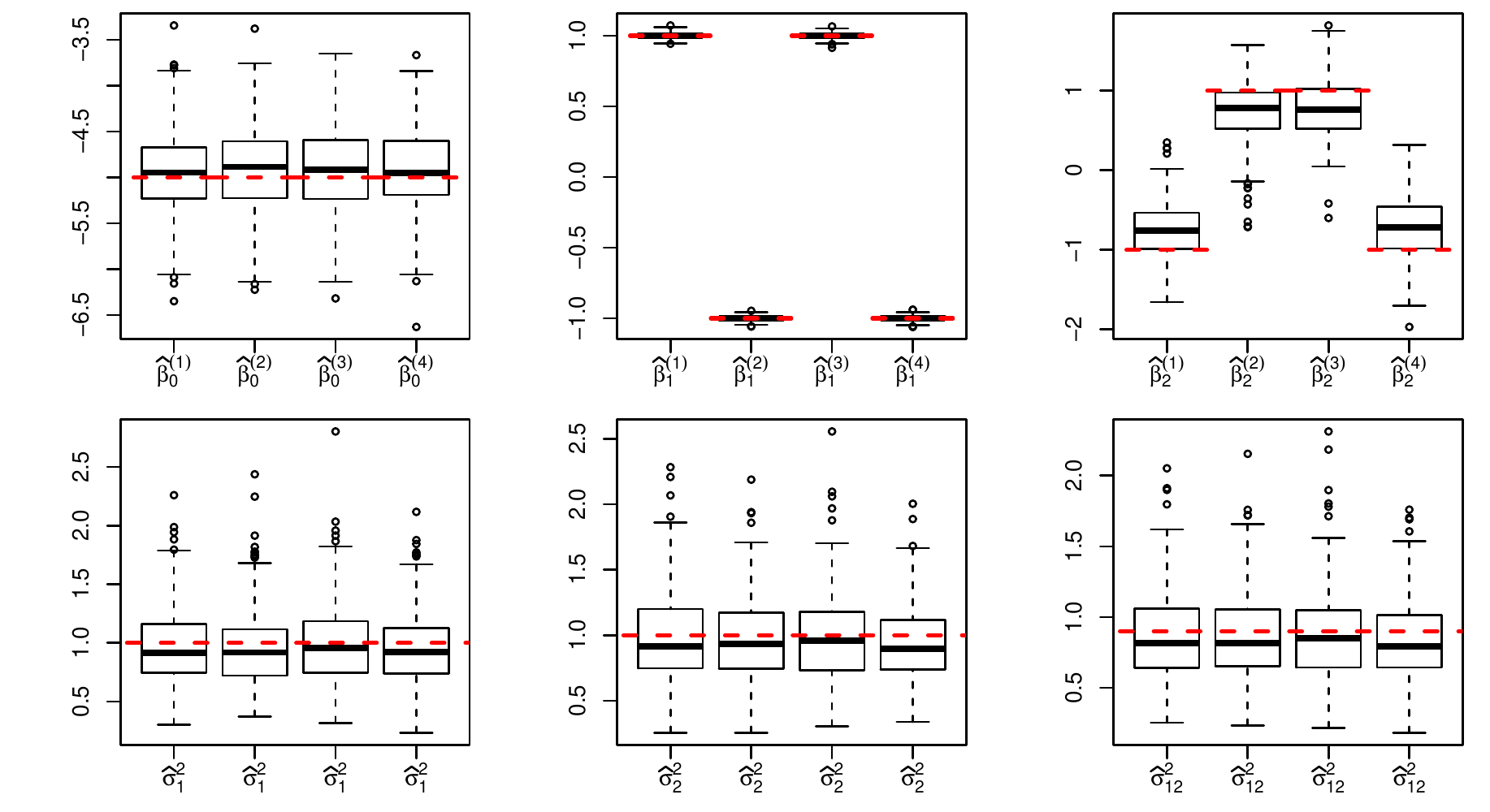}
\caption{\label{fig: boxplots_parameters} Fixed parameter estimates obtained from a simulation study with $S = 250$ replications for every scenario $k = 1,\dots,4$. The true values of $\mathbold{\beta}$ are given by $\mathbold{\beta}^{(1)} = (-5, 1, -1)^\top, \mathbold{\beta}^{(2)} = (-5, 1, -1)^\top, \mathbold{\beta}^{(3)} = (-5, 1, -1)^\top, \mathbold{\beta}^{(4)} = (-5, 1, -1)^\top$ and the components of $\mathbf{\Sigma}$ are specified by \eqref{eq: Sigma_simulation} in all cases. In the boxplots, the true values are marked by a dashed line.}
\end{figure}

We rerun the data-generating process as well as the data-simulating process $S = 250$ times and fit the model with four different parameter vectors $\mathbold{\beta}^{(1)},\dots,\mathbold{\beta}^{(4)}$. In Figure \ref{fig: boxplots_parameters} we summarize the resulting fixed parameter estimates. For the sake of compareability with the true value of $\mathbf{\Sigma}$, we indicate the estimate $\mathbf{\Sigma}$ which we would get when calculating the sum in \eqref{eq: updata_Sigma} leaving out the index $w$.

We can see that in general, the estimates for the intercept and the time-dependent parameter vary around the true value with a low variance with respect to the mean, especially for the time-dependent effect. The estimates for the dyadic-specific effect can be recognized to be rather biased towards zero and exhibiting a higher variance. However, in most of the simulations the sign of the estimate $\widehat{\beta}_2^{(k)}$ in scenario $k = 1,\dots,4$ is correct. The estimates of the fixed variance and covariance components $\sigma_{1}^2, \sigma_{2}^2$ and $\sigma_{12}^2$ are nearly unbiased but tend to show a higher variance. This could be remedied when more stations are included in the study. Overall, we see for the fixed parameters a promising performance.

\section{Discussion}
\label{sec: discussion}

Despite just few information on temporal network flows in a bike-sharing network given by station feeds only, our model is able to capture a large part of the effects that determine the network flow. Especially time-dependent effects and station-dependent (random) effects can be estimated properly with little uncertainty. Over a longer period the estimates of incoming and outgoing bikes could be shown to be rather precise for both real data application and simulated data. In the simulation study it was also shown that dyadic-specific effects are somewhat biased towards zero and that the standard errors of the estimates are much higher than for time-dependent effects. Finally, checking and incorporating possible overdispersion would be an advisable option though this will lead to a much more complex model. 

Our model can easily be applied to any network with integer count temporal network flows if the filling levels $\mathcal{C}_{i,t}$ of each node $i$ are known on an equidistant time grid. In \eqref{eq: mu_i_dot_t} and \eqref{eq: mu_dot_j_t} we implicitly define a routing matrix $A$ (see e.g. \citeauthor{medina2002traffic}, \citeyear{medina2002traffic}) which in our case specifies that every of the $N^2$ possible edges can have a positive weight. By changing the index sets in the sums of \eqref{eq: mu_i_dot_t} and \eqref{eq: mu_dot_j_t}, arbitrary routing matrices can be implied, e.g. setting the weights of loops to zero corresponds to the usual case when conducting traffic matrix estimation.

\section*{Acknowledgements}
We want to thank ZAMG (Vienna) for providing the weather data as well as  Michael Sedlmair (University of Stuttgart) and  Michael Oppermann (University of British Columbia) for providing the station feed data. Special thanks go to the elite graduate program Data Science at LMU Munich and the Munich Center for Machine Learning (MCML) for funding. Furthermore, the project was supported by the European Cooperation in Science and Technology [COST Action CA15109 (COSTNET)].

\FloatBarrier

\bibliographystyle{Chicago}
\bibliography{BikeSharing_arXiv}

\newpage

\pagenumbering{Roman}
\appendix

\section{Derivatives of the Log-Likelihood}

\label{app: skellam}

Since the series included in \eqref{eq: modified bessel function} converges absolutely, one can show that
\begin{align*}
\frac{\partial}{\partial \theta_1} I_d \left(2 \sqrt{\theta_1 \theta_2}\right) &= \frac{d}{2\theta_1} I_d \left(2 \sqrt{\theta_1 \theta_2} \right) + \sqrt{\frac{\theta_2}{\theta_1 }} I_{d+1}\left(2 \sqrt{\theta_1 \theta_2}\right), \\
\frac{\partial}{\partial \theta_2} I_d \left(2 \sqrt{\theta_1 \theta_2}\right) &= \frac{d}{2\theta_2} I_d \left(2 \sqrt{\theta_1 \theta_2} \right) + \sqrt{\frac{\theta_1}{\theta_2 }} I_{d+1}\left(2 \sqrt{\theta_1 \theta_2}\right).
\end{align*}
Using this relation, the partial derivatives of the log-likelihood contribution $l_D = l_D(\theta_1, \theta_2; d)$ with respect to $\theta_1$ and $\theta_2$ are given by 
\begin{align*}
\frac{\partial l_D}{\partial \theta_1} = -1 + \frac{d}{\theta_1} + \sqrt{\frac{\theta_2}{\theta_1}} \frac{I_{d+1}\left(2 \sqrt{\theta_1 \theta_2}\right)}{I_{d}\left(2 \sqrt{\theta_1 \theta_2}\right)}, \quad \frac{\partial l_D}{\partial \theta_2} = -1 + \sqrt{\frac{\theta_1}{\theta_2}}  \frac{I_{d+1}\left(2 \sqrt{\theta_1 \theta_2}\right)}{I_{d}\left(2 \sqrt{\theta_1 \theta_2}\right)}.
\end{align*}

The second-order partial derivatives of $l_D$ with respect to $\theta_1$ and $\theta_2$ are given by 
\begin{align*}
\frac{\partial^2 l_D}{\partial \theta_1^2} &= -\frac{d}{\theta_1^2} + \frac{\theta_2}{\theta_1} \left[ \frac{ I_{d+2}(2\sqrt{\theta_1 \theta_2}) I_{d}(2\sqrt{\theta_1 \theta_2}) - I_{d+1}(2\sqrt{\theta_1 \theta_2})^2}{I_{d}(2\sqrt{\theta_1 \theta_2})^2} \right], \\
\frac{\partial^2 l_D}{\partial \theta_2^2} &=  \frac{\theta_1}{\theta_2} \left[ \frac{ I_{d+2}(2\sqrt{\theta_1 \theta_2}) I_{d}(2\sqrt{\theta_1 \theta_2}) - I_{d+1}(2\sqrt{\theta_1 \theta_2})^2}{I_{d}(2\sqrt{\theta_1 \theta_2})^2} \right]
\intertext{and}
\frac{\partial^2 l_D}{\partial \theta_2 \partial \theta_1} = \frac{\partial^2 l_D}{\partial \theta_1 \partial \theta_2} &= \frac{1}{\sqrt{\theta_1 \theta_2}} \frac{I_{d+1}(\sqrt{2\theta_1 \theta_2})}{I_{d}(\sqrt{2\theta_1 \theta_2})} \notag \\ &+  \left[ \frac{ I_{d+2}(2\sqrt{\theta_1 \theta_2}) I_{d}(2\sqrt{\theta_1 \theta_2}) - I_{d+1}(2\sqrt{\theta_1 \theta_2})^2}{I_{d}(2\sqrt{\theta_1 \theta_2})^2} \right]. 
\end{align*}

The direct calculations of $\log I_d(\theta)$ and $I_{d+k}(\theta)/I_d(\theta)$ for $k = 1,2$ are not possible if $I_d(\theta)$ does not converge numerically. If this is the case, we use approximations of the Modified Bessel functions proposed by \cite{amos1974computation}. They show that for $\theta, d \geq 0$ it holds that
\begin{equation}
\label{eq: approx ratio bessel}
0 \leq \frac{\theta}{d+\frac{1}{2}+\sqrt{\theta^2 + (d + \frac{3}{2})^2}} \leq \frac{I_{d+1}(\theta)}{I_d(\theta)} \leq \frac{\theta}{d+\frac{1}{2}+\sqrt{\theta^2 + (d + \frac{1}{2})^2}} \leq 1.
\end{equation} 
We approximate $I_{d+1}(\theta)/I_d(\theta)$ as the mean value of the lower and the upper bound. The fraction $I_{d+2}(\theta)/I_d(\theta)$ can be approximated applying \eqref{eq: approx ratio bessel} twice. Furthermore, for $0 \leq \widetilde{\theta} \leq \theta$ the value of $I_d(\theta)$ can be bounded by $L(\theta, \widetilde{\theta}, d) \leq I_d(\theta) \leq U(\theta, \widetilde{\theta}, d)$, where
\begin{align}
\label{eq: L bound}
L(\theta, \widetilde{\theta}, d) &= \left( \frac{\theta}{\widetilde{\theta}} \right)^d I_d(\widetilde{\theta}) \exp \left( \frac{\theta^2 - \widetilde{\theta}^2}{\sqrt{\theta^2 + a^2} + \sqrt{\widetilde{\theta}^2 + a^2}} \right) \left( \frac{b+ \sqrt{\widetilde{\theta}^2 + a^2}}{b+ \sqrt{\theta^2 + a^2}} \right)^{d + \frac{1}{2}} \\
\label{eq: U bound}
U(\theta, \widetilde{\theta}, d) &= \left( \frac{\theta}{\widetilde{\theta}} \right)^d I_d(\widetilde{\theta}) \exp \left( \frac{\theta^2 - \widetilde{\theta}^2}{\sqrt{\theta^2 + b^2} + \sqrt{\widetilde{\theta}^2 + b^2}} \right) \left( \frac{b+ \sqrt{\widetilde{\theta}^2 + b^2}}{b+ \sqrt{\theta^2 + b^2}} \right)^{d + \frac{1}{2}}
\end{align}
with 
\begin{equation*}
a = d+ \frac{3}{2}, \quad b = d + \frac{1}{2}.
\end{equation*}
If $\widetilde{\theta} = \theta$, the boundaries are equal. Hence, we choose $\widetilde{\theta}$ as the maximum value such that $I_d(\widetilde{\theta})$ converges numerically. An approximate value of $\log I_d(\theta)$ is then given by the mean of the logarithm of the boundaries \eqref{eq: L bound} and \eqref{eq: U bound}.

\section{Details on the Maximization Algorithm}

\label{app: algorithm}

In every loop of Algorithm \ref{alg: EM} we need to maximize the penalized log-likelihood 
\begin{equation*}
l_P(\mathbold{\theta}) =  \sum_{i \in \lbrace 1, \dots, N, w \rbrace} \sum_{t \in \mathcal{T}} l_D(\mathbold{\theta}; d_{i, t}) - \frac{1}{2} \sum_{m=1}^M \lambda_m \mathbold{\gamma}_m^\top \mathbold{K}_m \mathbold{\gamma}_m - \frac{1}{2} \sum_{i \in \lbrace 1, \dots, N, w \rbrace} \mathbold{u}_i^\top \mathbf{\Sigma}^{-1} \mathbold{u}_i
\end{equation*}
with respect to $\mathbold{\theta} = \left(\mathbold{\beta}^\top, \mathbold{\gamma}_1^\top, \dots, \mathbold{\gamma}_M^\top, \mathbold{u}_1^\top, \dots, \mathbold{u}_N^\top, \mathbold{u}_w^\top\right)^\top$ if $\mathbf{\Sigma}$ and $\mathbold{\lambda}$ are assumed to be fixed. There is no analytical expression of the Fisher-information involving Skellam distributed random variables known yet and it appears that the observed Fisher-Information is not positive definite for most of the choices of $\mathbold{\theta}$ such that a Fisher-Scoring algorithm usually does not converge to a local optimum. Therefore, we use a quasi-Newton algorithm  already implemented in \textbf{R} to maximize the penalized log-likelihood. More precisely, we use the function \texttt{optim} with the method "BFGS". This algorithm does not need second order derivatives and is hence suitable for our problem. For more details on the BFGS algorithm, see \cite{wright1999numerical}. Hence, we need a representation of the derivative of $l_P$ with respect to $\mathbold{\theta}$. Denoting 
\begin{align*}
\theta_1 =  \theta_1(\mathbold{\theta}; i, t) &= \sum_{j \in \lbrace 1,\dots,N,w \rbrace} \exp\left( \eta(\mathbold{z}_{ji,t}) +  u_j^{\text{out}}  +u_i^{\text{in}}  \right), \\
\theta_2 =  \theta_2(\mathbold{\theta}; i, t) &= \sum_{j \in \lbrace1,\dots,N,w \rbrace} \exp\left( \eta(\mathbold{z}_{ij,t}) +  u_i^{\text{out}}  +u_j^{\text{in}}  \right)
\end{align*}
the components of the penalized score function $s_P$ are given by 
\begin{align*}
\mathbold{s}_{P_\beta} &= \frac{\partial l_P(\mathbold{\theta})}{\partial \mathbold{\beta}} = \sum_{i} \sum_{t}  \frac{\partial l_D}{\partial \theta_1} \frac{\partial \theta_1}{\partial \mathbold{\beta} } + \frac{\partial l_D}{\partial \theta_2} \frac{\partial \theta_2}{\partial \mathbold{\beta} } , \\
\mathbold{s}_{P_{\gamma_m}} &= \frac{\partial l_P(\mathbold{\theta})}{\partial \mathbold{\gamma}_m} = \sum_{i} \sum_{t} \frac{\partial l_D}{\partial \theta_1} \frac{\partial \theta_1}{\partial \mathbold{\gamma}_m} + \frac{\partial l_D}{\partial \theta_2} \frac{\partial \theta_2}{\partial \mathbold{\gamma}_m} - \lambda_m \mathbold{K}_m \mathbold{\gamma}_m, \\
\mathbold{s}_{P_{\mathbf{u}_i}} &= \frac{\partial l_P(\mathbold{\theta})}{\partial \mathbold{u}_i} = \sum_{i} \sum_{t}  \frac{\partial l_D}{\partial \theta_1} \frac{\partial \theta_1}{\partial \mathbold{u}_i } + \frac{\partial l_D}{\partial \theta_2} \frac{\partial \theta_2}{\partial \mathbold{u}_i }  - \mathbf{\Sigma}^{-1} \mathbold{u}_i.
\end{align*}

Even though we don't need the observed Fisher-Information to maximize $l_P$, we need its inverse in order to perform the updates of $\widehat{\mathbf{\Sigma}}$ and $\widehat{\mathbold{\lambda}}$ according to \eqref{eq: updata_Sigma} and \eqref{eq: update_lambda} as well as to compute the standard errors which are stated in Table \ref{tab: linear_effects}. Applying the chain rule and the product rule we have for example
\begin{align*}
\mathbold{F}_{P_{\beta\beta}}^\text{obs} = -\frac{\partial^2 l_P(\mathbold{\theta})}{\partial \mathbold{\beta} \partial \mathbold{\beta}^\top} &= -\sum_{i} \sum_{t} \left( \left[ \frac{\partial^2 l_D}{\partial \theta_1^2} \frac{\partial \theta_1}{\partial \beta} + \frac{\partial^2 l_D}{\partial \theta_2 \partial \theta_1} \frac{\partial \theta_2}{\partial \mathbold{\beta}} \right] \frac{\partial \theta_1}{\partial \mathbold{\beta}^\top} + \frac{\partial l_D}{\partial \theta_1} \frac{\partial^2 \theta_1}{\partial \mathbold{\beta} \partial \mathbold{\beta}^\top} \right) \\
&-\sum_{i} \sum_{t}  \left( \left[ \frac{\partial^2 l_D}{\partial \theta_1 \partial \theta_2} \frac{\partial \theta_1}{\partial \mathbold{\beta}} + \frac{\partial^2 l_D}{\partial \theta_2^2} \frac{\partial \theta_2}{\partial \mathbold{\beta}} \right] \frac{\partial \theta_2}{\partial \mathbold{\beta}^\top} + \frac{\partial l_D}{\partial \theta_2} \frac{\partial^2 \theta_2}{\partial \mathbold{\beta} \partial \mathbold{\beta}^\top} \right).
\end{align*}
Again, the remaining components of the observed Fisher-Information $\mathbold{F}_P^\text{obs}$ can be calculated in a similar manner. Finally we denote
\begin{align*}
\widehat{\mathbold{V}} = (\mathbold{F}_P^{\text{obs}})^{-1}(\widehat{\mathbold{\theta}}) = \begin{pmatrix}\widehat{\mathbold{V}}_{\beta \beta} & \widehat{\mathbold{V}}_{\beta \gamma_1} & \hdots & \widehat{\mathbold{V}}_{\beta \gamma_M} & \widehat{\mathbold{V}}_{\beta u_1} & \hdots & \widehat{\mathbold{V}}_{\beta u_N} & \widehat{\mathbold{\mathbold{V}}}_{\beta u_w} \\
\widehat{\mathbold{V}}_{\beta \gamma_1} & \widehat{\mathbold{V}}_{\gamma_1 \gamma_1} & \hdots & \widehat{\mathbold{V}}_{\gamma_M \gamma_1} & & & & \widehat{\mathbold{\mathbold{V}}}_{\gamma_1 u_w}  \\
\vdots & \vdots & \ddots & \vdots & & & & \vdots \\
\widehat{\mathbold{V}}_{\beta \gamma_M} & \widehat{\mathbold{V}}_{\gamma_1 \gamma_M} & \hdots & \widehat{\mathbold{V}}_{\gamma_M \gamma_M} &  & & & \widehat{\mathbold{\mathbold{V}}}_{\gamma_M u_w} \\
\widehat{\mathbold{V}}_{u_1 \beta} & & & &  \widehat{\mathbold{V}}_{u_1 u_1} & \hdots & \widehat{\mathbold{V}}_{u_1 u_N} & \widehat{\mathbold{\mathbold{V}}}_{u_1 u_w} \\
\vdots & & & & \vdots & \ddots & \vdots & \vdots \\
\widehat{\mathbold{V}}_{u_N \beta} &   & &   & \widehat{\mathbold{V}}_{u_N u_1} & \hdots & \widehat{\mathbold{V}}_{u_N u_N} & \widehat{\mathbold{\mathbold{V}}}_{u_N u_w} \\
\widehat{\mathbold{\mathbold{V}}}_{u_w \beta} & \widehat{\mathbold{\mathbold{V}}}_{u_w \gamma_1} & \hdots & \widehat{\mathbold{\mathbold{V}}}_{u_w \gamma_M} &  \widehat{\mathbold{\mathbold{V}}}_{u_w u_1} & \hdots & \widehat{\mathbold{\mathbold{V}}}_{u_w u_N} & \widehat{\mathbold{\mathbold{V}}}_{u_w u_w}
 \end{pmatrix}.
\end{align*}
as the estimated variance-covariance matrix of $\widehat{\mathbold{\theta}}$.

\section{Model Evaluation -- Continued}

\label{app: model evaluation}

\begin{figure}[t]
\centering
\includegraphics[width = 0.49\textwidth]{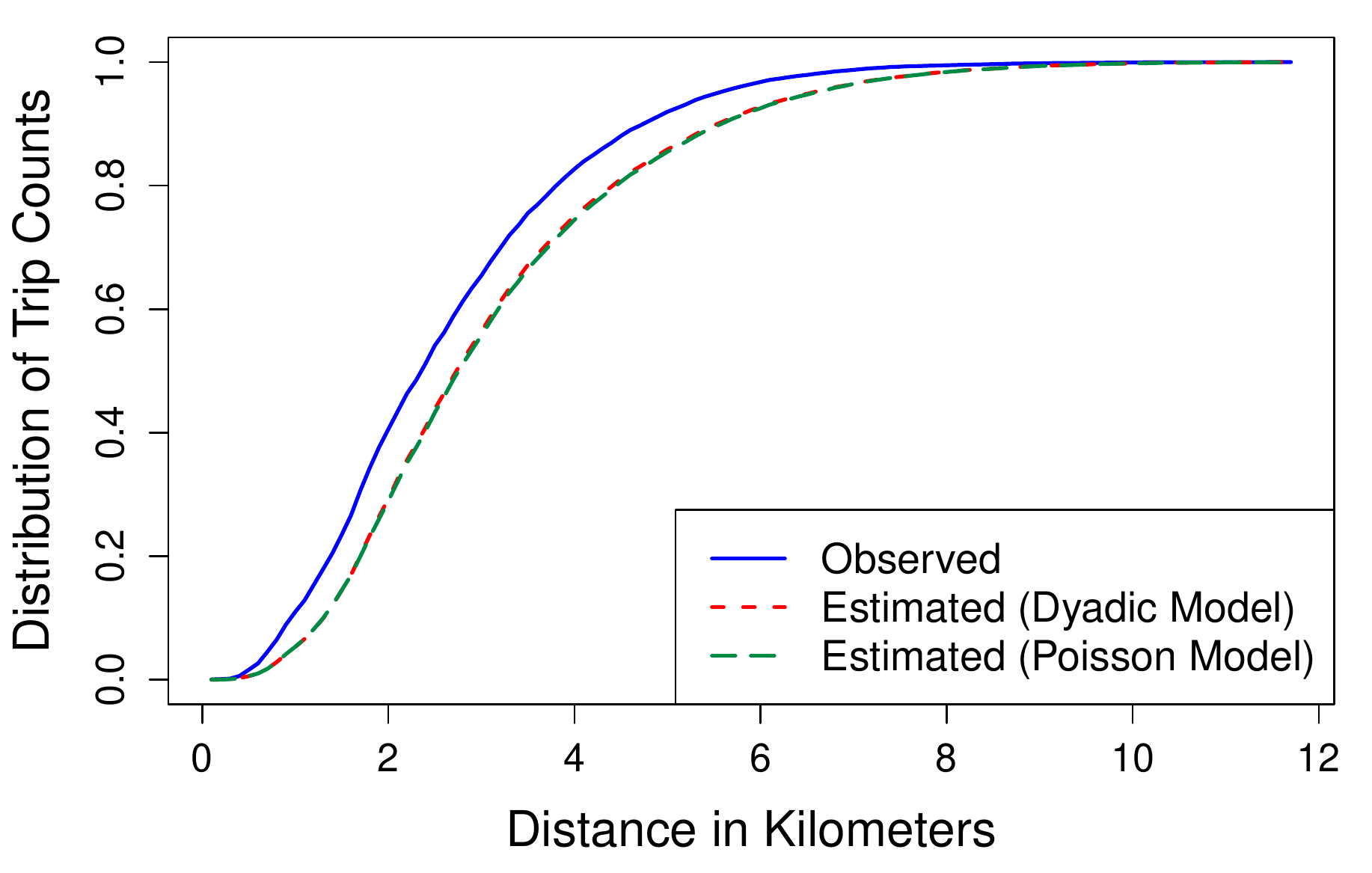}
\includegraphics[width = 0.49\textwidth]{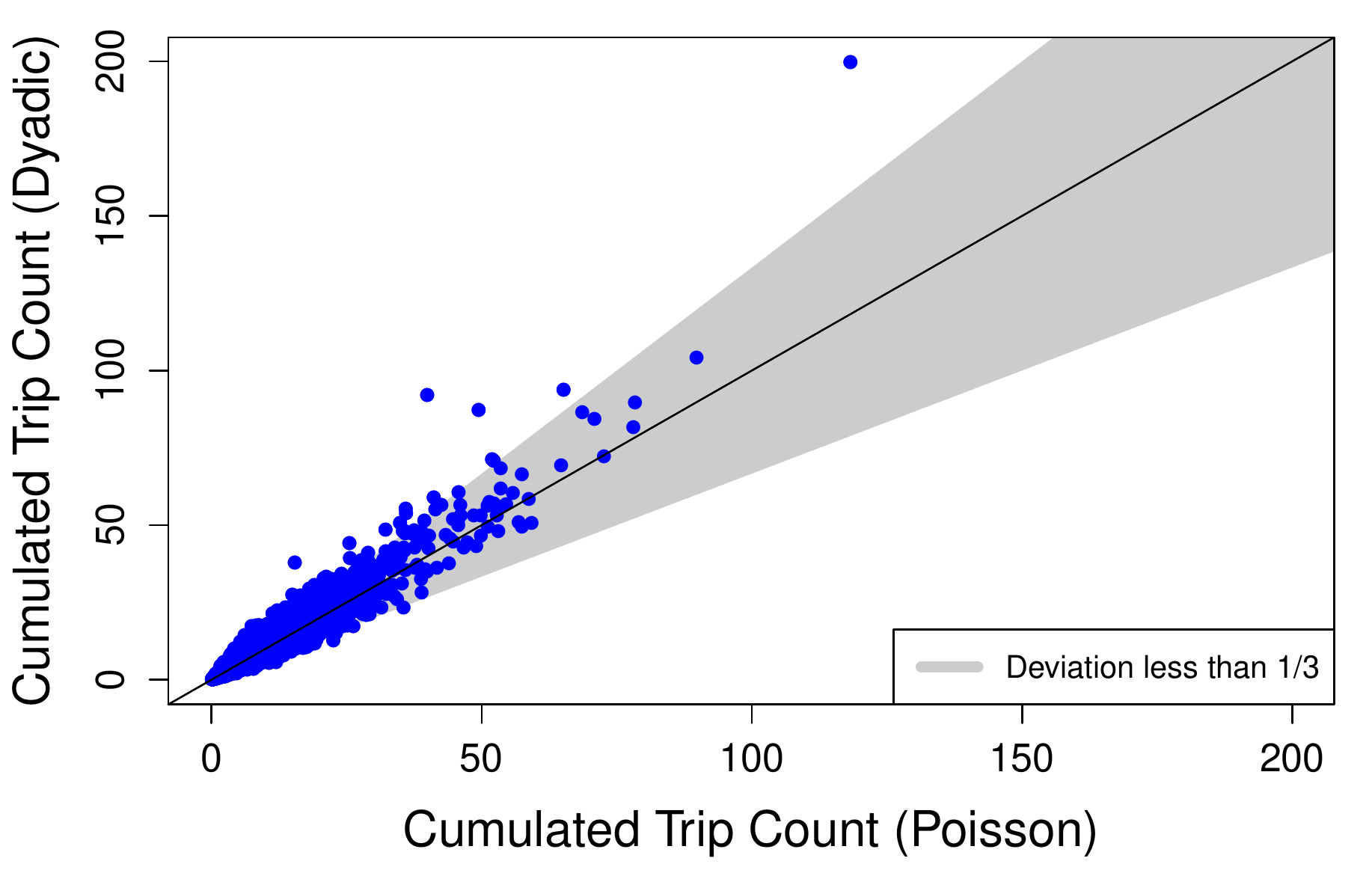}
\caption{\label{fig: dyad model} Left panel: Observed vs. estimated distribution of trip counts by distance covered in the network, loops are left out; Right panel: Estimates of cumulated trip counts for every connection in the network}
\end{figure}

In this section we further assess the performance of the dyadic network flow prediction model \eqref{eq: param_1} where we also compare the results to those of the corresponding Poisson model. In the left panel of Figure \ref{fig: dyad model} we contrast the observed with the estimated cumulative distribution of trip counts by distance covered in the network. The tendency of all three curves is the same, but for any distance $\texttt{dist}$, the count of estimated trips covering at most $\texttt{dist}$ is always smaller than actually observed. Furthermore, the Poisson model`s curve is very close to the dyadic model's curve.

The right panel of Figure \ref{fig: dyad model} opposes the estimates of the cumulated trip counts over the set of time points $\mathcal{T}$
\begin{equation*}
\widehat{\mu}_{ij,\sbullet}^\text{Poisson} = \sum_{t \in \mathcal{T}} \widehat{\mu}_{ij,t}^\text{Poisson}, \quad \widehat{\mu}_{ij,\sbullet}^\text{dyad} = \sum_{t \in \mathcal{T}} \widehat{\mu}_{ij,t}^\text{dyad}
\end{equation*}
estimated with the Poisson model to the dyadic model's estimates. Here, we can also see that there is a general concordance. For the majority of the most frequent used connections in the network the relative deviations of the estimates are less than 1/3. 

\begin{figure}[t]
\centering
\includegraphics[width = \textwidth]{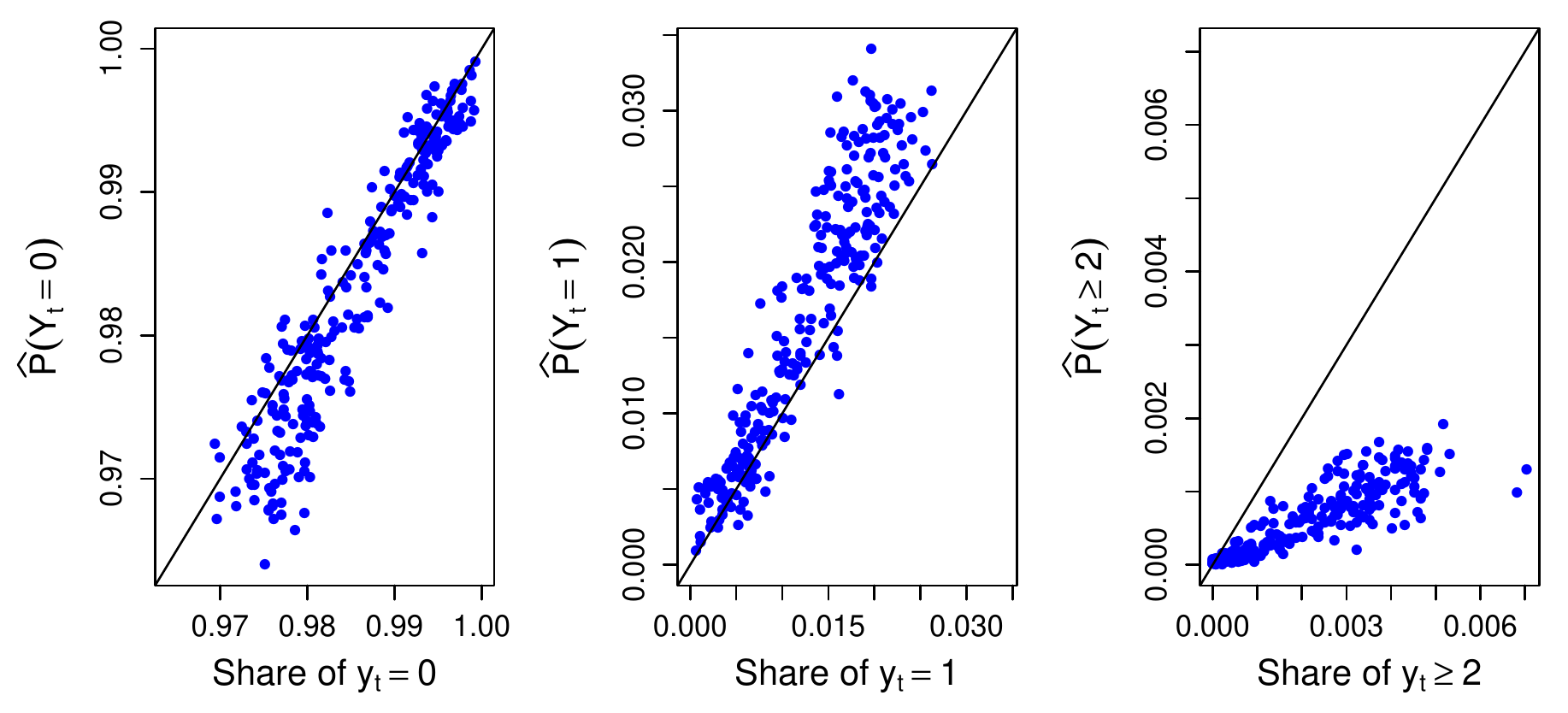}
\caption{\label{fig: probabilities} Observed vs. estimated probabilities of trip counts for $t \in \mathcal{T}$}
\end{figure}

In Figure \ref{fig: probabilities} we compare for every $t \in \mathcal{T}$ the observed shares of trip counts being either zero, one or greater than one with the corresponding estimates which are given by
\begin{align*}
\widehat{\mathbb{P}}(Y_{ t} = 0) &= \frac{1}{N^2} \sum_{i,j = 1}^N \exp(-\widehat{\mu}_{ij,t}), \quad \widehat{\mathbb{P}}(Y_{t} = 1) = \frac{1}{N^2} \sum_{i,j = 1}^N  \widehat{\mu}_{ij,t}\exp(-\widehat{\mu}_{ij,t}), \\
\widehat{\mathbb{P}}(Y_{t} \geq 2) &= \frac{1}{N^2} \sum_{i,j = 1}^N \left[1- \exp(-\widehat{\mu}_{ij,t}) - \widehat{\mu}_{ij,t}\exp(-\widehat{\mu}_{ij,t})\right]
\end{align*}
due to the Poisson assumption. The typical large count of zero-trip counts in bike-sharing systems can be captured very well, the maximal deviation is only 1.3\%. The percentages of trip counts being equal to one are rather overestimated with an average deviation of 24.9\%. The share of trip counts larger than one is distinctly underestimated for all $t$. Thus, it can be concluded that there is overdispersion in the data which a Poisson model is not able to capture.

\end{document}